# Accumulation tests for FDR control in ordered hypothesis testing

Ang Li and Rina Foygel Barber

06.10.16

**Abstract**

Multiple testing problems arising in modern scientific applications can involve simultaneously testing thousands or even millions of hypotheses, with relatively few true signals. In this paper, we consider the multiple testing problem where prior information is available (for instance, from an earlier study under different experimental conditions), that can allow us to test the hypotheses as a ranked list in order to increase the number of discoveries. Given an ordered list of $n$ hypotheses, the aim is to select a data-dependent cutoff $k$ and declare the first $k$ hypotheses to be statistically significant while bounding the false discovery rate (FDR). Generalizing several existing methods, we develop a family of "accumulation tests" to choose a cutoff $k$ that adapts to the amount of signal at the top of the ranked list. We introduce a new method in this family, the HingeExp method, which offers higher power to detect true signals compared to existing techniques. Our theoretical results prove that these methods control a modified FDR on finite samples, and characterize the power of the methods in the family. We apply the tests to simulated data, including a high-dimensional model selection problem for linear regression. We also compare accumulation tests to existing methods for multiple testing on a real data problem of identifying differential gene expression over a dosage gradient.

## 1 Introduction

In many modern applications of statistics, the availability of high-dimensional data sets allows simultaneous testing of a large list of potential hypotheses. This can lead to the well-known issue of multiple testing—if each hypothesis is tested with a significance threshold $\alpha$, e.g. $\alpha = 0.05$, then in the scenario where the number of true signals among the $n$ hypotheses is small, we can expect $\sim \alpha \cdot n$ false discoveries. Many methods have been developed to handle this issue, including the Benjamini-Hochberg procedure [3] where the threshold $\alpha$ is chosen in a way that adapts to the amount of signal in the data, to control the proportion of false discoveries. Historically, many methods have been developed that treat the $n$ hypotheses interchangeably, while in practice we may often have additional information—for instance, a group structure on the set of hypotheses (where we believe that the true signals are likely to cluster in these groups), a hierarchical structure (for instance, in a regression problem, we might believe that an interaction effect between variables $X_j$ and $X_k$ should only be present if $X_j$ and $X_k$ each show marginal effects on the response as well), or an ordering or partial ordering (where prior information leads us to believe that some hypotheses are more likely to contain a true signal than others, or where we want to test hypotheses in a certain order to respect the structure of the problem). In this work, we treat this last case.

To describe the problem more precisely, suppose that we are interested in testing $n$ hypotheses, denoted $H_1, \ldots, H_n$, and experimental data has yielded individual p-values for each of these hypotheses, which we write as $p_1, \ldots, p_n$. For a concrete example, we might be searching for a genetic cause for some particular disease, in which case we might test a hypothesis $H_i$ that the $i$th SNP in our experiment is associated with the disease, where the index $i = 1, \ldots, n$ labels each of the SNPs being tested. The $i$th p-value $p_i$ would be a measure of the association between the $i$th SNP and the disease, and we assume a null distribution $p_i \sim \mathsf{Uniform}[0, 1]$ whenever the $i$th SNP has no association with the disease.

In many practical settings, the experiment that we would like to analyze has been carried out in the context of existing information from previous studies. At the same time, often this prior information cannot simply be treated as additional data in the statistical analysis. For instance, in the example above where $p_i$ is a p-value testing the association between SNP $i$ and some phenotype of interest, we might have prior information available from earlier experiments that may have:

- studied a different but related disease;
- tested a different population of patients;
- used a different experimental protocol for genotyping the individuals;
- defined disease status differently, or measured a real-valued phenotype differently; or
- produced data that we believe may be unreliable.

In any of these scenarios, the data from the previous study cannot simply be integrated with our new experimental data, without violating the integrity of the statistical analysis. However, this prior information is extremely valuable and can give us some power to detect signals in a very high-dimensional setting (e.g. $n$ in the millions or more, with very few true signals).

While the scenarios described above can correspond to many different forms of prior information about the $n$ hypotheses being tested, in this work we focus on the specific problem of testing the hypotheses when the only prior information is a ranking of the list. Before performing the experiment, we use prior information to generate a ranked list of hypotheses $H_1, H_2, \ldots, H_n$, where $H_1$ is the hypothesis that we believe is most likely to correspond to a true signal, while $H_n$ is the one believed to be least likely. After gathering new data, we then wish to test these hypotheses while taking this ordering into account.

**Application to high-dimensional regression** In addition to situations where prior information, or data from related experiments, may provide a ranked list, this type of setting is also applicable to other statistical problems. As a key example, consider the problem of inference for sparse regression, where a response $y$ depends on some sparse subset of many possible features $X_1, \ldots, X_p$. A sparse linear model for this data states that $y = X\beta + \epsilon$, where $\beta \in \mathbb{R}^p$ is a sparse vector of coefficients while $\epsilon$ contains the (zero-mean) noise in the response variable $y$. When the sample size $n$ is lower than the number of features $p$, classical methods for performing inference on the coefficients of $\beta$, such as testing hypotheses of the form $\beta_j = 0$ or finding confidence intervals for the coefficients $\beta_j$, cannot be applied as the linear model is not identifiable (i.e. $X^\top X$ is rank-deficient). In the recent literature, many approaches have been proposed for this inference problem, including a recent line of work by Taylor et al. [37] that, when paired with the Lasso (penalized regression) or with a forward stepwise selection procedure, calculates p-values for each feature in the order that they are selected. This provides a list of p-values with an inherent ordering, and therefore is an example of the ordered hypothesis testing problem we consider here.

**Outline** The remainder of this paper is organized as follows. Section 2 formally introduces the sequential testing problem considered here, gives background on several existing methods, and develops our family of methods that generalizes these existing works. In this section we also summarize some related problems and existing methods, and discuss our method in the context of the surrounding literature. Section 3 presents the theoretical results of this paper, including results for FDR control and for the power of the method in both finite-sample and asymptotic settings. These theoretical results are proved in Section 4, with some proofs deferred to the appendix. Section 5 presents experiments on simulated data to validate our theoretical results and provide an empirical comparison of various choices of the method within the general family. In Section 6 we adapt our method to a dosage-response problem, where for a large set of genes, we would like to determine which genes respond to a particular drug and, for the responsive genes, what is the minimal dosage that induces a response—this section includes an experiment on real data demonstrating the effectiveness of our method in this setting. We conclude with a discussion of our findings and future directions in Section 7.

Code implementing the accumulation test methods in R [29], along with code to reproduce the simulated data experiment and the gene dosage data experiment, is available online.[1]

# 2 Problem and method

We begin by formally defining the problem that we consider here. Let $\mathcal{H}_0 \subseteq \{1, \ldots, n\}$ be a fixed set (the "null hypotheses"), and let $p_1, \ldots, p_n \in [0, 1]$ be random variables, such that $p_i \stackrel{\text{iid}}{\sim} \mathsf{Unif}[0, 1]$ for all $i \in \mathcal{H}_0$, and furthermore the null p-values are independent of the non-null p-values. Our method will construct a cutoff point $\widehat{k}$ that is adaptive to the data—formally, this cutoff is a function mapping the observed p-values $(p_1, \ldots, p_n)$ to some $\widehat{k} \in \{0, \ldots, n\}$. This cutoff $\widehat{k}$ is the output of our procedure, and should be interpreted as labeling the first $\widehat{k}$ hypotheses, i.e. $H_1, \ldots, H_{\widehat{k}}$, as "discoveries" (to use the terminology of hypothesis testing, we reject hypotheses $H_1, \ldots, H_{\widehat{k}}$ and do not reject $H_{\widehat{k}+1}, \ldots, H_n$).

Ideally, we would like to choose $\widehat{k}$ so that the selected list $H_1, \ldots, H_{\widehat{k}}$ contains only true signals and the remaining hypotheses $H_{\widehat{k}+1}, \ldots, H_n$ are all null. However, this may not be possible because our initial ranking may be imperfect—the ranked list $H_1, \ldots, H_n$ may contain signals and nulls interspersed with each other, meaning that there is no threshold that perfectly separates the signals from the nulls. However, the ranking is nonetheless informative if the signals are indeed concentrated towards the top of the ranked list, and we select $\widehat{k}$ with the goal of detecting as many signals as possible without too many false positives. To quantify this, we define the false discovery proportion (FDP) cumulatively along the list:

$$\mathsf{FDP}(k) = \frac{\mathsf{FalsePos}(k)}{k} \ ,$$

where $\mathsf{FalsePos}(k) = \#\{i \leq k : i \in \mathcal{H}_0\}$ is the number of false positives among the first $k$ hypotheses. In other words, $\mathsf{FDP}(k)$ gives the proportion of false positives (i.e. null hypotheses) among the first $k$ hypotheses in the list, i.e. $H_1, \ldots, H_k$. To agree with the definition of false discovery rate used in the literature, we define $\mathsf{FDP}(0) := 0$ to cover the case that no discoveries are made; more formally, we can write $\mathsf{FDP}(k) = \frac{\mathsf{FalsePos}(k)}{\max\{1,k\}}$ to cover both cases $k = 0$ and $k \neq 0$.

[1] `http://www.stat.uchicago.edu/~rina/accumulationtests.html`

For ease of notation, we will omit this more precise definition and will write $\frac{\mathsf{FalsePos}(k)}{k}$ with the understanding that $\frac{0}{0}$ is treated as 0.

Selecting a threshold $\widehat{k}$ involves a tradeoff: we would like a high $\widehat{k}$ to ensure that as many true signals as possible are captured in the selected list $H_1, \dots, H_{\widehat{k}}$, but at the same time the proportion of false positives will generally increase farther down the list, since the signals will be concentrated towards the top of the ranking. In particular, we would like to bound the false discovery rate (FDR) [3], defined as the expectation of the FDP (where the expectation is taken with respect to the distribution of the p-values):

$$\text{False discovery rate} = \mathbb{E}\left[\mathsf{FDP}(\widehat{k})\right] = \mathbb{E}\left[\frac{\mathsf{FalsePos}(\widehat{k})}{\widehat{k}}\right] .$$

**A note on the ranking** In the setting described in the Introduction, where prior experience or preexisting data allows us to rank the hypotheses from most to least likely, this ranking is reflected in the indexing of the list. That is, before gathering data for the current experiment, we find the hypothesis that we deem to be most likely and label it $H_1$, then the next most likely hypothesis is labeled $H_2$, etc. Implicit in the assumption above is the idea that this ranking takes place before the p-values are generated, that is, the p-values are independent from the process of ranking the hypotheses. For instance, we cannot use a data set to rank the hypotheses and then use the same data set to calculate p-values.

**Existing methods** Suppose that we would like to select a cutoff $\widehat{k}$ that is as large as possible, while bounding the FDR at some prespecified level $\alpha$ (e.g. $\alpha = 0.1$). Two approaches for the ordered hypothesis testing problem have been proposed recently in the literature. First, G'Sell et al. [22] propose the ForwardStop method:

$$\widehat{k}_{\mathsf{ForwardStop}} = \max\left\{k \in \{1, \dots, n\} : \frac{1}{k}\sum_{i=1}^{k} \log\left(\frac{1}{1-p_i}\right) \leq \alpha\right\} , \tag{1}$$

with the convention that if this set is empty, we set $\widehat{k}_{\mathsf{ForwardStop}} = 0$ and make no rejections. To understand this method, consider a single null hypothesis $H_i$. Since $p_i \sim \mathsf{Uniform}[0, 1]$ by assumption, it is true that $\log\left(\frac{1}{1-p_i}\right) \sim \mathsf{Exp}(1)$ and in particular, $\mathbb{E}\left[\log\left(\frac{1}{1-p_i}\right)\right] = 1$. Then, for any fixed potential cutoff point $k$,

$$\mathbb{E}\left[\sum_{i=1}^{k} \log\left(\frac{1}{1-p_i}\right)\right] \geq \mathbb{E}\left[\sum_{i \leq k, i \in \mathcal{H}_0} \log\left(\frac{1}{1-p_i}\right)\right] = \mathsf{FalsePos}(k) .$$

This implies that, for any fixed $k$, the cumulative sum $\frac{1}{k}\sum_{i=1}^{k} \log\left(\frac{1}{1-p_i}\right)$ is, in expectation, at least as large as $\mathsf{FDP}(k)$. Therefore, the cutoff $\widehat{k}_{\mathsf{ForwardStop}}$ is the last time when the *estimated* FDP lies at or below the predetermined level $\alpha$. G'Sell et al. [22, Theorem 1] prove that this procedure controls FDR at the level $\alpha$, that is,

$$\mathbb{E}\left[\mathsf{FDP}(\widehat{k}_{\mathsf{ForwardStop}})\right] \leq \alpha .$$

A second existing method uses a similar summation to estimate the FDP, but with a discrete step function rule rather than continuous measure: given a parameter $C > 1$, the Sequential Step-up

Procedure (SeqStep) of Barber and Candès [1] sets

$$\widehat{k}_{\mathsf{SeqStep}(C)} = \max\left\{ k \in \{1,\dots,n\} : \frac{1}{k}\sum_{i=1}^{k} C\cdot \mathbb{1}\{p_i > 1-1/C\} \leq \alpha \right\} . \tag{2}$$

As for the ForwardStop procedure [22], we see that this cumulative summation serves to (over)estimate the FDP: for any null hypothesis $H_i$, clearly $\mathbb{E}\left[C\cdot \mathbb{1}\{p_i > 1-1/C\}\right] = 1$ and so

$$\mathbb{E}\left[\sum_{i=1}^{k} C\cdot \mathbb{1}\{p_i > 1-1/C\}\right] \geq \mathbb{E}\left[\sum_{i\leq k, i\in\mathcal{H}_0} C\cdot \mathbb{1}\{p_i > 1-1/C\}\right] = \mathsf{FalsePos}(k) .$$

Barber and Candès [1, Theorem 3] prove that this procedure controls a modified form of the FDR:

$$\mathbb{E}\left[\frac{\#\{i \leq \widehat{k}_{\mathsf{SeqStep}(C)} : i \in \mathcal{H}_0\}}{C/\alpha + \widehat{k}_{\mathsf{SeqStep}(C)}}\right] \leq \alpha . \tag{3}$$

For exact FDR control, the same paper proposes a slightly more conservative variant, the SeqStep+ method [1], defined by

$$\widehat{k}_{\mathsf{SeqStep}_+(C)} = \max\left\{ k \in \{1,\dots,n\} : \frac{1}{1+k}\left(C + \sum_{i=1}^{k} C\cdot \mathbb{1}\{p_i > 1-1/C\}\right) \leq \alpha \right\} , \tag{4}$$

with the guarantee that for any $C > 1$,

$$\mathbb{E}\left[\mathsf{FDP}(\widehat{k}_{\mathsf{SeqStep}_+(C)})\right] \leq \alpha .$$

**A general family of methods** Examining the two procedures described above, G'Sell et al. [22]'s ForwardStop procedure and Barber and Candès [1]'s SeqStep procedures, we see that the two have a common structure: they each compute an (over)estimate of the false discovery proportion at each point in the path, $\mathsf{FDP}(k)$, via two different choices of cumulative sums. We now generalize these two procedures with a broader family:

**Definition 1** (Accumulation test for ordered hypotheses)**.** Suppose we are given a ranked list of $n$ hypotheses $H_1,\dots,H_n$ with corresponding p-values $p_1,\dots,p_n \in [0,1]$. Fix any function $\mathsf{h} : [0,1] \mapsto [0,\infty)$ that satisfies $\int_{t=0}^{1} \mathsf{h}(t)\ \mathsf{d}t = 1$; we call $\mathsf{h}$ the "accumulation function". Define the estimated FDP at each cutoff $k \in \{1,\dots,n\}$ as

$$\widehat{\mathsf{FDP}}_{\mathsf{h}}(k) = \frac{\sum_{i=1}^{k} \mathsf{h}(p_i)}{k} ,$$

and then select the adaptive cutoff

$$\widehat{k}_{\mathsf{h}} = \max\left\{ k \in \{1,\dots,n\} : \widehat{\mathsf{FDP}}_{\mathsf{h}}(k) \leq \alpha \right\} ,$$

where $\alpha \in (0,1)$ is a prespecified target FDR level. (We use the convention that $\widehat{k}_{\mathsf{h}} = 0$ if this set is empty.) We then reject the hypotheses $H_1,\dots,H_{\widehat{k}_{\mathsf{h}}}$ and do not reject $H_{\widehat{k}_{\mathsf{h}+1}},\dots,H_n$.

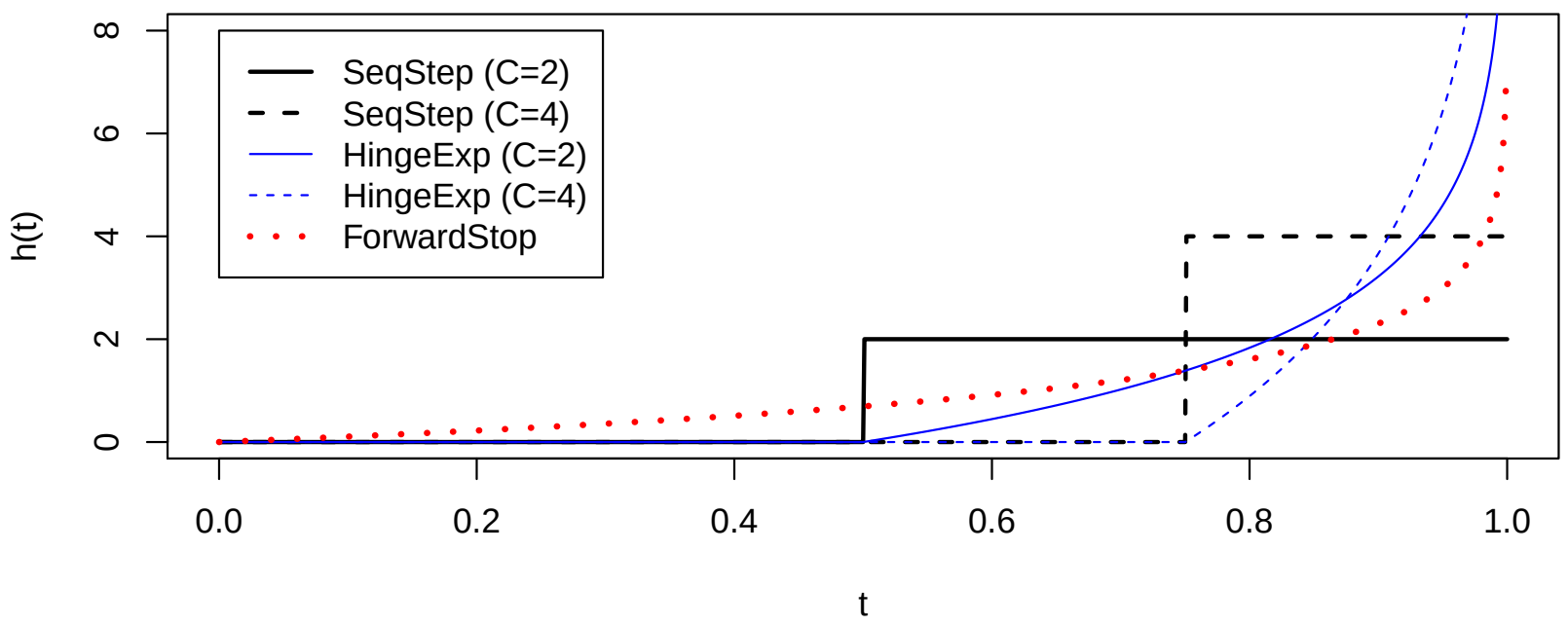

*Figure 1: Illustration of several choices of the accumulation function* $\mathsf{h} : [0, 1] \to [0, \infty)$.

Note that our notation implicitly treats the desired FDR bound $\alpha$ as fixed, while $\mathsf{h}$ appears in the subscript to emphasize that we may choose between many candidate accumulation functions $\mathsf{h}$. In particular, choosing the accumultation function $\mathsf{h}_{\mathsf{ForwardStop}}(t) = \log\left(\frac{1}{1-t}\right)$ yields G'Sell et al. [22]'s ForwardStop procedure, while $\mathsf{h}_{\mathsf{SeqStep}}(t) = C \cdot \mathbb{1}\{t > 1 - 1/C\}$ yields Barber and Candès [1]'s SeqStep procedure (for any parameter $C > 1$). We will also study an additional choice for the accumulation function $\mathsf{h}$, the "HingeExp" function,

$$\mathsf{h}_{\mathsf{HingeExp}}(t) = \begin{cases} C \cdot \log\left(\frac{1}{C\cdot(1-t)}\right), & \text{for } t > 1 - 1/C, \\ 0, & \text{for } t \leq 1 - 1/C. \end{cases} \tag{5}$$

This function combines the ideas of the step function used in SeqStep with the ForwardStop method; Figure 1 gives a visual comparison of these methods. The name "HingeExp" arises from the "hinge point" of the function at $t = 1 - 1/C$ (similar to the hinge loss function in machine learning), combined with the observation that, for a null p-value $p_i \sim \mathsf{Unif}[0, 1]$, we have $\mathsf{h}_{\mathsf{HingeExp}}(p_i)$ distributed as $C$ times an $\mathsf{Exp}(1)$ random variable with probability $1/C$ (and equal to zero otherwise). We will see that the HingeExp method offers superior empirical performance compared with existing options, in many settings.

For any accumulation function $\mathsf{h}$, as before the data-dependent cutoff $\widehat{k}_{\mathsf{h}}$ should intuitively control the false discovery rate at the desired level. This is because, for any fixed $k$, the sum $\sum_{i=1}^{k} \mathsf{h}(p_i)$ is the estimated "accumulation" of false positives by the time we have reached the $k$th position in the list:

$$\mathbb{E}\left[\sum_{i=1}^{k} \mathsf{h}(p_i)\right] \geq \mathbb{E}\left[\sum_{i \leq k, i \in \mathcal{H}_0} \mathsf{h}(p_i)\right] = \mathsf{FalsePos}(k) \ ,$$

where the last step holds because $\mathbb{E}\left[\mathsf{h}(p_i)\right] = 1$ for $i \in \mathcal{H}_0$ due to the requirement $\int_{t=0}^{1} \mathsf{h}(t) \ \mathsf{d}t = 1$. Therefore, for each $k$, the estimated false discovery proportion $\widehat{\mathsf{FDP}}_{\mathsf{h}}(k)$ is an overestimate of the actual FDP:

$$\mathbb{E}\left[\widehat{\mathsf{FDP}}_{\mathsf{h}}(k)\right] = \frac{\mathbb{E}\left[\sum_{i=1}^{k} \mathsf{h}(p_i)\right]}{k} \geq \frac{\mathsf{FalsePos}(k)}{k} = \mathsf{FDP}(k) \ . \tag{6}$$

**Preview of theoretical results** Our first main result, Theorem 1, proves that, for any choice of the accumulation function $\mathsf{h}$, the method described above satisfies a bound on modified FDR similar to the bound (3) obtained in [1], with no further assumptions other than the ones described above. In the simplest case, for a desired false discover rate bound $\alpha$, and for any bounded accumulation function $\mathsf{h} : [0,1] \to [0,C]$, we have

$$\mathbb{E}\left[\frac{\mathsf{FalsePos}(\widehat{k}_{\mathsf{h}})}{C/\alpha + \widehat{k}_{\mathsf{h}}}\right] \leq \alpha \ .$$

This result is given in Theorem 1, along with generalizations to other formulations of the test and to unbounded accumulation functions. Of course, the quantity whose expectation is bounded, is slightly different from the FDP due to the added term $C/\alpha$ in the denominator. Intuitively, this should make little difference when $\widehat{k}_{\mathsf{h}}$ is large, and indeed in Theorem 2 we prove that the accumulation test controls the FDP asymptotically.

Since we are free to choose any accumulation function $\mathsf{h}$, we naturally would like to know how to choose this function to maximize our power for detecting signals. Intuitively, an accumultation function $\mathsf{h}$ will have good power if $\mathbb{E}\left[\mathsf{h}(p_i)\right]$ is as low as possible for the non-null p-values, since these expectations control the extent of the overestimation of FDP (see (6)). Our theoretical results for this setting show that lower expectations $\mathbb{E}\left[\mathsf{h}(p_i)\right]$ for $i \notin \mathcal{H}_0$ leads to higher power in an asymptotic setting (Theorem 3). Minimizing $\mathbb{E}\left[\mathsf{h}(p_i)\right]$ over the non-nulls $i \notin \mathcal{H}_0$ remains an open question, although somewhat surprisingly, if we restrict our attention to bounded accumulation functions only, we find that the the step functions used in the SeqStep procedure [1] achieve the best (lowest) expectations $\mathbb{E}\left[\mathsf{h}(p_i)\right]$ for non-nulls $i \notin \mathcal{H}_0$ (Lemma 2).

## 2.1 Related work

As discussed above, several existing methods treat the ordered hypothesis testing problem. Our method generalizes the ForwardStop procedure by G'Sell et al. [22] and the SeqStop procedure by Barber and Candès [1] for controlling FDR in ordered setting. Another procedure known to us is the $\alpha$-investing by Foster and Stine [18], which controls the the ratio $\frac{\mathbb{E}[V]}{\mathbb{E}[R]+1}$ (where $V$ = number of false positives and $R$ = total number of discoveries), a criterion weaker than the FDR. It allows users to incorporate prior knowledge such as ordering and improve the power. However, $\alpha$-investing shows lower power than ForwardStop in simulations carried out by G'Sell et al. [22].

Ordered testing procedures have been shown to provide FDR control in regression models. Taylor et al. [37] derived post-selection hypothesis tests at each step of the forward stepwise and LARS procedures. These tests yield an ordered list of p-values, corresponding to a nested sequence of models. With the ordered testing procedures, these sequential p-values can be transformed into a model selection procedure with FDR guarantee ([22]), which we explore in Section 5.

While research on ordered multiple testing is limited, there is rich literature on the general multiple testing and FDR control. Here we discuss three widely used methods: the Benjamini-Hochberg (BH) procedure [3], Storey [34]'s modification of the BH procedure, and the empirical Bayes method (Efron et al. [17], Efron and Tibshirani [16]). The BH procedure for FDR control works for any configuration of null and non-null hypotheses, given that the p-values corresponding to true nulls are independent. In the dependent case, Benjamini and Yekutieli [5] showed that the BH method still controls FDR if the joint distribution of statistics (or p-values) is PRDS on the set of true nulls (PRDS stands for "positive regression dependence on a subset", e.g. Gaussian variables with totally positive covariance matrix is PRDS), or if the BH procedure is conducted with the FDR level one $\log(n)$-th of the target FDR level, where $n$ is the total number of hypotheses. The BH procedure has been widely applied to examine large scale datasets, including microarray gene expression data ([30], [13]), brain fMRI ([20], [24]), etc. Storey [34]'s procedure replaces $n$ in the BH procedure with an estimate of the total number of nulls, which may be substantially smaller than $n$. Therefore, it is less conservative and has greater power than BH procedure ([34], [35]). The empirical Bayes method [17] calculates the Bayesian FDR as the posterior probability of null conditioned on rejection, from the estimates of prior probabilities and densities of nulls and non-nulls. Efron et al. [17] and Efron and Tibshirani [16] also proposed local FDR, the probability of rejecting a null in a subset of the rejection region, and showed its application in the analysis of breast cancer microarray data.

In multiple testing, when hypotheses share a hierarchical, or group dependence structure, this information can be utilized to improve the power of testing, and the interpretability of results. The hierarchical testing procedure by Benjamini and Yekutieli [6] incorporates known hierarchical dependence between hypotheses by arranging them on a tree. At each step, hypotheses within the same family (i.e. with the same immediate parents) are tested simultaneously, and for the significant cases, the process keeps moving forward. This procedure can target the control of FDR at full-tree level, at a certain depth, or among the outer-nodes. The Group BH procedure by Hu et al. [26] groups the hypotheses based on known dependence information (such as genes within the same biological pathways, or sharing the same phenotypes). The proportion of non-nulls in each group is then estimated, and the original p-values are reweighted to emphasize the groups with higher estimated proportions of non-nulls before the final BH step. Other studies or approaches that fall into this category are the pooled and separate analysis ([14]), the decision theoretical approach (Cai and Sun [8]), and the selection-adjusted procedure (Benjamini and Bogomolov [2]). In the brain fMRI application, Heller et al. [24] grouped voxel units into clusters using previous correlation data, and applied the BH method at the cluster level. Their approach enjoys greater interpretability, as well as increased power.

Besides hierarchical and grouped testing, there are several approaches in the literature to incorporate prior information into multiple testing. One approach is by weighting the hypotheses with prior information ([4], [21], [31], [26]). Another approach, mostly applied in microarray analysis, uses prior information to exclude non-informative genes before the final selection step of significant ones, which enhances the power ([7], [23], [27], [36]). Also worth noticing is the Bayesian mixture model approach to include previous knowledge in genome-wide linkage studies and association studies ([28], [19]). Recently, Du and Zhang [12] introduced a single-index modulated (SIM) procedure, which assumes the availability of a bivariate p-value $(p_1, p_2)$ (where $p_1$ is the p-value from prior information, and $p_2$ is the main p-value reflecting curent information), and project it into a single p-value combining $p_1$ and $p_2$ in some optimal direction for the final analysis. With prior information, this approach could improve the power significantly while maintaining the control of FDR.

Scott and Berger [33] explored a Bayesian hierarchical approach for multiple testing. The posterior probabilities, including the probability that hypothesis $i$ is null given the data, are inferred through importance sampling. They discussed the choice of prior distributions on model parameters. This approach has been applied in disease mapping [9], and abnormal corporate performance identification [32].

# 3 Theoretical results

In this section we develop and prove results on the FDR control properties, and the power, of the family of accumulation test methods. In Section 3.1 we prove that accumulation tests control the FDR (or modified FDR) in the finite sample and asymptotic settings. In Section 3.2 we examine the power of the accumulation tests in finite sample and asymptotic settings.

## 3.1 False discovery rate control

### 3.1.1 Finite sample setting

We begin with concrete finite-sample results to bound the false discovery rate of the family of accumulation tests. Recall that, when a threshold $k \in \{1, \dots, n\}$ is selected, we are interested in bounding the false discovery proportion

$$\mathsf{FDP}(k) = \frac{\mathsf{FalsePos}(k)}{k} = \frac{\#\{i \leq k : i \in \mathcal{H}_0\}}{k}$$

(with the convention that $\mathsf{FDP}(0) = 0$). We will also consider a modified form of the FDP, introduced in Barber and Candès [1] for the SeqStep method, given by

$$\mathsf{mFDP}_c(k) = \frac{\mathsf{FalsePos}(k)}{c + k} = \frac{\#\{i \leq k : i \in \mathcal{H}_0\}}{c + k} .$$

Of course, when $c$ is a constant while $k$ is large, the modified FDP is nearly identical to the FDP.

Our main result shows that the accumulation test controls the modified FDP. Furthermore, a slightly more conservative test (defined in the theorem) controls the original FDP.

**Theorem 1.** *Let* $\mathsf{h} : [0, 1] \to [0, \infty)$ *be any function with* $\int_{t=0}^{1} \mathsf{h}(t)\ \mathsf{d}t = 1$*, and let* $\alpha \in (0, 1)$ *be some prespecified target FDR level. Fix any* $C > 0$*. Define*

$$\widehat{k}_{\mathsf{h}} = \max\left\{k \in \{1, \dots, n\} : \frac{1}{k}\sum_{i=1}^{k} \mathsf{h}(p_i) \leq \alpha\right\} , \tag{7}$$

*with the convention that* $\widehat{k}_{\mathsf{h}} = 0$ *if this set is empty, and define*

$$\widehat{k}_{\mathsf{h}}^{+C} = \max\left\{k \in \{1, \dots, n\} : \frac{1}{1 + k}\left(C + \sum_{i=1}^{k} \mathsf{h}(p_i)\right) \leq \alpha\right\} , \tag{8}$$

*with the same convention. Then, in the special case that* $\max_{0 \leq t \leq 1} \mathsf{h}(t) \leq C$*, we have*

$$\mathbb{E}\left[\mathsf{mFDP}_{C/\alpha}(\widehat{k}_{\mathsf{h}})\right] \leq \alpha \quad \textit{and} \quad \mathbb{E}\left[\mathsf{FDP}(\widehat{k}_{\mathsf{h}}^{+C})\right] \leq \alpha . \tag{9}$$

*In the general case, with no restriction on the range of* $\mathsf{h}$*, we have*

$$\mathbb{E}\left[\mathsf{mFDP}_{C/\alpha}(\widehat{k}_{\mathsf{h}})\right] \leq \frac{\alpha}{\int_{t=0}^{1} \mathsf{h}(t) \wedge C\ \mathsf{d}t} \quad \textit{and} \quad \mathbb{E}\left[\mathsf{FDP}(\widehat{k}_{\mathsf{h}}^{+C})\right] \leq \frac{\alpha}{\int_{t=0}^{1} \mathsf{h}(t) \wedge C\ \mathsf{d}t} , \tag{10}$$

*where we use the notation* $a \wedge b := \min\{a, b\}$*.*

We also give a result specifically for the HingeExp function, which gives a tighter bound than that guaranteed by Theorem 1:

**Lemma 1.** *Let* $\mathsf{h}(t) = C \cdot \log\left(\frac{1}{C(1-t)}\right) \cdot \mathbb{1}_{t > 1 - 1/C}$*, i.e. the HingeExp function with parameter* $C$*. Then, under the same definitions and assumptions as Theorem 1,*

$$\mathbb{E}\left[\mathsf{mFDP}_{2C/\alpha}(\widehat{k}_{\mathsf{h}})\right] \leq \alpha .$$

**Comparison to existing results** As discussed in Section 2, the accumulation test contains two existing procedures as special cases: SeqStep [1] with the step function $\mathsf{h}(t) = C \cdot \mathbb{1}\{t > 1 - 1/C\}$, and ForwardStop [22] with $\mathsf{h}(t) = \log\left(\frac{1}{1-t}\right)$. Furthermore, the slightly altered accumulation test given in (8) contains as a special case the SeqStep+ procedure [1], again with $\mathsf{h}(t) = C \cdot \mathbb{1}\{t > 1 - 1/C\}$. For the special cases of SeqStep and SeqStep+, our main result, Theorem 1, obtains the same guarantee on modified and original FDP as proved in Barber and Candès [1, Theorem 3]. For the special case of the ForwardStop procedure, the results obtained in Theorem 1 and Lemma 1 are somewhat weaker than the guarantee given in G'Sell et al. [22, Theorem 1], which proves that

$$\mathbb{E}\left[\mathsf{FDP}(\widehat{k}_{\mathsf{h}})\right] \leq \alpha \ ,$$

that is, a guarantee on the FDP rather than the modified FDP. As we will see in Theorem 2, however, asymptotically the same result is obtained.

**A conjecture** Based on the results of Theorem 1 and Lemma 1, we conjecture the following generalization:

**Conjecture 1.** *Let* $\mathsf{h} : [0,1] \to [0,\infty)$ *be any accumulation function satisfying* $\int_{t=0}^{1} \mathsf{h}(t)\ \mathsf{d}t = 1$ *and* $\int_{t=0}^{1} \left[\mathsf{h}(t)\right]^2\ \mathsf{d}t \leq C$*, for some* $C \geq 1$*. Under the same definitions and assumptions as Theorem 1,*

$$\mathbb{E}\left[\mathsf{mFDP}_{C/\alpha}(\widehat{k}_{\mathsf{h}})\right] \leq \alpha \ .$$

Note that, if true, this conjecture would replicate the results of Theorem 1 for bounded functions and of Lemma 1 for the HingeExp function, and would strengthen the results of Theorem 1 for unbounded functions.

### 3.1.2 Asymptotic setting

In our finite-sample result (Theorem 1), we proved that the accumulation test controls a modified form of the FDP, which as discussed earlier, is nearly equal to the original FDP as long as the number of rejections $\widehat{k}_{\mathsf{h}}$ is large. Next, we show that the accumulation test controls FDR asymptotically as long as the number of rejections $\widehat{k}_{\mathsf{h}}$ tends to infinity.

**Theorem 2.** *Let* $\mathsf{h} : [0,1] \to [0,\infty)$ *be any function with* $\int_{t=0}^{1} \mathsf{h}(t)\ \mathsf{d}t = 1$*, and let* $\alpha \in (0,1)$ *be some prespecified target FDR level. Consider a sequence of ordered hypothesis testing problems, with* $n = 1, 2, \dots$*. Suppose that there exists a sequence* $m_n \in \mathbb{N}$ *with* $m_n \to \infty$ *such that*

$$\mathbb{P}\left\{\widehat{k}_{\mathsf{h}} < m_n\right\} \to 0 \text{ as } n \to \infty \ .$$

*Then*

$$\lim_{n\to\infty} \mathbb{E}\left[\mathsf{FDP}(\widehat{k}_{\mathsf{h}})\right] \leq \alpha \ .$$

## 3.2 Power

Up to this point, our discussion and theoretical analysis of accumulation tests has focused on controlling the false discovery rate. Of course, in practice we are interested in balancing the goals of reducing false positives (Type I error) while increasing the number of true discoveries (power). Formally, we will define the power of a cutoff $k \in \{1, \dots, n\}$ as the proportion of non-nulls that are discovered by the test,

$$\mathsf{Power}(k) = \frac{\#\{i \le k : i \notin \mathcal{H}_0\}}{\#\{i : i \notin \mathcal{H}_0\}} \ ,$$

and will aim to choose the accumulation function $\mathsf{h}$ to maximize the expected power of the accumulation test,

$$\mathbb{E}\left[\mathsf{Power}(\widehat{k}_{\mathsf{h}})\right] \ .$$

Recall that $\widehat{k}_{\mathsf{h}}$ is defined as the largest $k$ such that

$$\widehat{\mathsf{FDP}}_{\mathsf{h}}(k) = \frac{\sum_{i=1}^{k} \mathsf{h}(p_i)}{k} \le \alpha \ .$$

In other words, high power corresponds to a large value of $\widehat{k}$, which is possible only when the estimates $\widehat{\mathsf{FDP}}_{\mathsf{h}}(k)$ are low. Therefore, a good choice of the accumulation function $\mathsf{h}$ is one that allows for low estimates $\widehat{\mathsf{FDP}}_{\mathsf{h}}(k)$ of the false discovery proportion along the sequence of hypotheses. Consider the expectation of this estimate at any fixed $k$,

$$\begin{aligned}\mathbb{E}\left[\widehat{\mathsf{FDP}}_{\mathsf{h}}(k)\right] = \frac{\sum_{i=1}^{k} \mathbb{E}\left[\mathsf{h}(p_i)\right]}{k} &= \frac{\sum_{i \le k, i \in \mathcal{H}_0} \mathbb{E}\left[\mathsf{h}(p_i)\right]}{k} + \frac{\sum_{i \le k, i \notin \mathcal{H}_0} \mathbb{E}\left[\mathsf{h}(p_i)\right]}{k} \\ &= \mathsf{FDP}(k) + \frac{\sum_{i \le k, i \notin \mathcal{H}_0} \mathbb{E}\left[\mathsf{h}(p_i)\right]}{k} \ ,\end{aligned}$$

where as before the last step holds because we require $\int_{t=0}^{1} \mathsf{h}(t) \ \mathsf{d}t = 1$ and so $\mathbb{E}\left[\mathsf{h}(p_i)\right] = 1$ for $i \in \mathcal{H}_0$. Therefore, for a given choice of $\mathsf{h}$, we see that $\widehat{\mathsf{FDP}}_{\mathsf{h}}(k)$ is an overestimate of $\mathsf{FDP}(k)$, with bias given by

$$\frac{\sum_{i \le k, i \notin \mathcal{H}_0} \mathbb{E}\left[\mathsf{h}(p_i)\right]}{k} \ .$$

Since power will increase if the estimated false discovery proportion $\widehat{\mathsf{FDP}}_{\mathsf{h}}(k)$ is small, we see that a good choice of accumulation function $\mathsf{h}$ is one that minimizes $\mathbb{E}\left[\mathsf{h}(p_i)\right]$ for non-nulls $i \notin \mathcal{H}_0$ (while, of course, satisfying the requirement $\mathbb{E}\left[\mathsf{h}(p_i)\right] = 1$ for $i \in \mathcal{H}_0$).

In the following section, we will examine how $\mathbb{E}\left[\mathsf{h}(p_i)\right]$ affects the power by studying an asymptotic scenario. We will find the power of any accumulation function $\mathsf{h}$ can be characterized exactly by its expected value for non-null p-values (Theorem 3). If all the non-null p-values follow a single distribution, $p_i \overset{\text{iid}}{\sim} \mathcal{D}$ for $i \notin \mathcal{H}_0$, we can think of these results as characterizing the power of an accumulation function $\mathsf{h}$ for testing the null hypothesis given by the uniform distribution against the alternate hypothesis given by the distribution $\mathcal{D}$.

Of course, in practice, we will not always know the distribution of the non-null p-values. In Section 3.2.2 we discuss the problem of determining a good choice of accumulation function $\mathsf{h}$ without prior knowledge of an alternate distribution.

Before we proceed, we recall the definition of a subexponential random variable:

$$X \text{ is subexponential with parameters } (\sigma^2, b) \text{ if } \mathbb{E}\left[e^{\theta(X - \mathbb{E}[X])}\right] \le e^{\theta^2\sigma^2/2} \text{ for all } |\theta| \le \frac{1}{b}. \quad (11)$$

Note that a $\sigma^2$-subgaussian random variable, such as a $N(0,\sigma^2)$ variable, is $(\sigma^2, 0)$-subexponential trivially, and so the subexponential condition is weaker than the subgaussian condition.

We will say that an accumulation function $\mathsf{h}$ is $(\sigma^2, b)$-subexponential with respect to the p-values if $\mathsf{h}(p_i)$ is $(\sigma^2, b)$-subexponential for each $i = 1, \dots, n$. In particular, note that the SeqStep, ForwardStop, and HingeExp functions are all subexponential (as long as the p-values are all dominated by the uniform distribution).

### 3.2.1 Power calculation in an asymptotic setting

In this section, we show that we can calculate the asymptotic power of the accumulation methods, and compare between them, under the assumption that the proportion of non-nulls along the list is converging to a fixed function $f(\cdot)$, where $f(\cdot)$ must satisfy some mild conditions. Specifically, we consider an asymptotic scenario where

$$\max_{k=0,\dots,n} \left| \frac{\#\{i \le k : i \notin \mathcal{H}_0\}}{k} - f\left(\frac{k}{n}\right) \right| \le \epsilon_n \ , \tag{12}$$

and $\epsilon_n \to 0$ as $n \to \infty$. We assume that $f : [0,1] \to [0,1]$ is differentiable and satisfies, for some constant $\delta > 0$ and for the fixed target FDR level $\alpha$,

$$\begin{cases} f'(t) \le 0 \text{ for all } t, \\ f'(t) \le -\delta \text{ for all } t \text{ such that } f(t) \ge 1-\alpha, \\ t \mapsto t \cdot f(t) \text{ is a nondecreasing function.} \end{cases} \tag{13}$$

In words, these conditions require that the proportion of true signals decreases as we move along the list; that the proportion decreases at a rate bounded away from zero during the initial portion of the list; and that the *number* of true signals must of course increase as we move along the list.

Then we have the following theorem:

**Theorem 3.** *Fix any $\sigma^2 > 0$ and $b \ge 0$, any target FDR level $\alpha \in [0,1]$, and any $\mu \in (0,1)$. Suppose that:*

- *The p-values are independent, with $p_i \overset{\text{iid}}{\sim} \mathsf{Unif}[0,1]$ for all $i \in \mathcal{H}_0$ and $p_i \overset{\text{iid}}{\sim} \mathcal{D}$ for all $i \notin \mathcal{H}_0$ for some distribution $\mathcal{D}$;*
- *The function $\mathsf{h} : [0,1] \mapsto [0,\infty)$ satisfies $\mathbb{E}_{p_i \sim \mathsf{Unif}[0,1]}[h(p_i)] = 1$ and $\mathbb{E}_{p_i \sim \mathcal{D}}[h(p_i)] = \mu$, and is $(\sigma^2, b)$-subexponential with respect to the p-values; and*
- *The function $f : [0,1] \to [0,1]$ satisfies assumptions (12) and (13).*

*Define*

$$T = \begin{cases} 0, & \text{if } \frac{1-\alpha}{1-\mu} \ge f(0), \\ f^{-1}\left(\frac{1-\alpha}{1-\mu}\right) \in (0,1), & \text{if } f(1) < \frac{1-\alpha}{1-\mu} < f(0), \\ 1, & \text{if } \frac{1-\alpha}{1-\mu} \le f(1). \end{cases}$$

*Then the power converges as*

$$\mathsf{Power}(\widehat{k}_{\mathsf{h}}) = \frac{\left|\{1,\dots,\widehat{k}_{\mathsf{h}}\} \backslash \mathcal{H}_0\right|}{N_1(n)} \to T \cdot \frac{f(T)}{f(1)} \ ,$$

*where specifically this denotes convergence in probability as $n \to \infty$.*

*Remark* 1. Note that, when holding $\alpha$ and $f(\cdot)$ fixed, $T$ is a nonincreasing function of $\mu = \mathbb{E}_{p_i \sim \mathcal{D}}\left[\mathsf{h}(p_i)\right]$. Therefore, the (asymptotic) power is a nonincreasing function of $\mu$ as well, since $T \mapsto T \cdot f(T)$ is nondecreasing. This means if $\mathbb{E}\left[\mathsf{h}_0(p_i)\right] \leq \mathbb{E}\left[\mathsf{h}(p_i)\right]$, for $i \notin \mathcal{H}_0$, the accumulation test using $\mathsf{h}_0(\cdot)$ is asymptotically more powerful than that using $\mathsf{h}(\cdot)$.

### 3.2.2 Choosing the accumulation function

While our earlier results prove control of the (modified) FDR for a broad class of accumulation functions, without any knowledge about the presence or absence of true signals, our understanding of the power of these methods does depend on the "alternate hypothesis", i.e. the distribution of the non-null p-values. In particular, Theorem 3 shows that the power of some specific accumulation function $\mathsf{h}$ can be viewed, asymptotically, as a simple function of this alternate hypothesis. However, without knowledge of this alternate hypothesis, how can we proceed—that is, how can we choose an effective $\mathsf{h}$?

Here we suggest two partial answers to this question. First, observe that our FDR results, Theorems 1 and 2, rely only on the *exchangeability* of the null p-values $\{p_i : i \in \mathcal{H}_0\}$, rather than requiring a strict i.i.d. assumption. Therefore, we can observe the *unordered* set of p-values, $\{p_i : i = 1, \dots, n\}$ before choosing the accumulation function $\mathsf{h}$, without negating the FDR control properties of the method. In particular, we may then use an empirical Bayes method (e.g. see Efron [15]) to estimate the distribution of the non-null $p_i$'s, which can then guide us in selecting $\mathsf{h}$.

Quite unexpectedly, in one special case it is possible to choose an optimal $\mathsf{h}$ without any knowledge of the alternate distribution: the case of bounded accumulation functions. The following result shows that the step function $\mathsf{h}(t) = C \cdot \mathbb{1}_{t > 1 - 1/C}$, which is used in the SeqStep method of [1], is the optimal $C$-bounded accumulation function under a very mild assumption on the distribution of non-null p-values:

$$\text{The non-null p-value } p_i \text{ has a density } f_i : [0,1] \to [0,\infty), \text{ where } f_i \text{ is a nonincreasing function.} \tag{14}$$

We say that $p_i$ satisfies the assumption (14) *strictly* if its density $f_i$ is a strictly decreasing function.

**Lemma 2.** *Consider any accumulation function* $\mathsf{h}$ *bounded by* $C \geq 1$*, that is,* $\mathsf{h} : [0,1] \to [0, C]$ *satisfies* $\int_{t=0}^{1} \mathsf{h}(t)\ \mathsf{d}t = 1$*. Let* $\mathsf{h}_0$ *be the step function with the same bound,* $\mathsf{h}_0(t) = C \cdot \mathbb{1}\{t > 1 - 1/C\}$*. Suppose that a non-null p-value* $p_i$ *satisfies the assumption* (14)*. Then*

$$\mathbb{E}\left[\mathsf{h}(p_i)\right] \geq \mathbb{E}\left[\mathsf{h}_0(p_i)\right] \ .$$

*Furthermore, the inequality is strict whenever* $p_i$ *satisfies* (14) *strictly, unless* $\mathsf{h}(t) = \mathsf{h}_0(t)$ *almost everywhere on* $t \in [0,1]$.

In other words, based on the discussion above, we expect the step function (the SeqStep method) to offer more power than any other accumulation function that maps to the same range, as long as the non-null p-values satisfy the assumption (14).

The assumption (14) is very natural, because we expect non-null p-values to give evidence against the null hypothesis, i.e. the distribution of a non-null $p_i$ should place more mass on low values (near 0) than on high values (near 1). As a specific example, suppose that we are performing a z-test on statistics $Z_i \overset{\perp\!\!\!\perp}{\sim} N(\mu_i, 1)$, where the $i$th null hypothesis is that $\mu_i = 0$. We therefore compute p-values with a two-tailed z-test, $p_i = 2(1 - \Phi(|Z_i|))$, where $\Phi$ is the standard normal cumulative distribution function. Then for any non-null $i$, with $Z_i \sim N(\mu_i, 1)$ for $\mu_i \neq 0$, $p_i$ follows the density

$$f_i(t) = \frac{e^{\mu_i \Phi^{-1}(1-t/2)} + e^{-\mu_i \Phi^{-1}(1-t/2)}}{2e^{-\mu_i^2/2}}$$

which is a strictly decreasing function of $t$ due to the fact that $x \mapsto e^x + e^{-x}$ is a strictly increasing function of $x$ when $x \geq 0$.

We expect that similar results may be possible under weaker assumptions on the accumulation function $\mathsf{h}$ (e.g. an assumption of bounded moments or subexponential tails), and leave this question to future work.

# 4 Proofs

In this section, we give the proofs of some of our main results: Theorems 1 and 2 for FDR control, and Theorem 3 for the power, with some details deferred to the Appendix. The proofs of Lemmas 1 and 2 are given in Appendix A.5 and Appendix A.3, respectively.

## 4.1 Proof of Theorem 1 (finite-sample FDR control)

The key ingredient for the proof of Theorem 1 is the following lemma:

**Lemma 3.** *Let $a_1, \dots, a_n \geq 0$ be any fixed thresholds, and let*

$$\widehat{k} = \max\left\{k \in \{1, \dots, n\} : \sum_{i=1}^{k} \mathsf{h}(p_i) \leq a_k\right\} , \tag{15}$$

*with the convention that $\widehat{k} = 0$ if this set is empty. Then*

$$\mathbb{E}\left[\frac{1 + \#\{i \leq \widehat{k} : i \in \mathcal{H}_0\}}{C + \sum_{i \leq \widehat{k}, i \in \mathcal{H}_0} \mathsf{h}(p_i)}\right] \leq \frac{1}{\int_{t=0}^{1} \mathsf{h}(t) \wedge C \,\mathsf{d}t} \, .$$

To understand the role of this result in proving Theorem 1, first note that the definitions of $\widehat{k}_{\mathsf{h}}$ and $\widehat{k}_{\mathsf{h}}^{+C}$, given in (7) and (8), can each be rewritten as a threshold criterion of the form (15) as given in Lemma 3.

Essentially, Lemma 3 shows that, at $k = \widehat{k}_{\mathsf{h}}$ (or at $k = \widehat{k}_{\mathsf{h}}^{+C}$), we have $\sum_{i=1}^{k} \mathsf{h}(p_i) \gtrapprox \#\{i \leq k : i \in \mathcal{H}_0\}$, and thus, this result guarantees that the estimated FDP, $\widehat{\mathsf{FDP}}_{\mathsf{h}}(k) = \frac{\sum_{i=1}^{k} \mathsf{h}(p_i)}{k}$, is a reliable (over)estimate of the actual FDP, $\mathsf{FDP}(k) = \frac{\#\{i \leq k : i \in \mathcal{H}_0\}}{k}$. Given this lemma, the proof of the bounds in Theorem 1 follows the arguments in Barber and Candès [1, Theorem 3]; we give details in Appendix A.1.

In order to prove Lemma 3, we use a result that treats the Bernoulli case specifically:

**Lemma 4** (Adapted from [1, Lemma 4])**.** *Let $B_1, \dots, B_n \in \{0, 1\}$ be independent, with $B_i \stackrel{\text{iid}}{\sim}$ $\mathsf{Bernoulli}(\rho)$ for all $i \in \mathcal{H}_0$. Let $\{\mathcal{F}_k\}_{i=1,\dots,n}$ be any filtration in reverse time (i.e. $\mathcal{F}_k \supseteq \mathcal{F}_{k+1}$) such*

*that*

$$B_i \in \mathcal{F}_k \textit{ for all } i \notin \mathcal{H}_0, \textit{ and for all } i > k \textit{ with } i \in \mathcal{H}_0, \tag{16}$$

$$\sum_{i \le k, i \in \mathcal{H}_0} B_i \in \mathcal{F}_k, \textit{ and} \tag{17}$$

$$\{B_i : i \le k, i \in \mathcal{H}_0\} \textit{ are exchangeable with respect to } \mathcal{F}_k, \tag{18}$$

*for all* $k = 1, \ldots, n$. *Then*

$$M_k = \frac{1 + \#\{i \le k : i \in \mathcal{H}_0\}}{1 + \sum_{i \le k, i \in \mathcal{H}_0} B_i}$$

*is a supermartingale with respect to* $\{\mathcal{F}_k\}$, *and* $\mathbb{E}[M_n] \le \frac{1}{\rho}$.

Equipped with this result for the special case of Bernoulli variables, we turn to the proof of our key lemma, where we construct a coupling between the general case and the Bernoulli case.

*Proof of Lemma 3.* First, recall that the null p-values, $p_i$ for $i \in \mathcal{H}_0$, are i.i.d. draws from $\mathsf{Unif}[0,1]$ and are independent from the non-null p-values. Therefore, we can treat the non-null p-values, $p_i$ for $i \notin \mathcal{H}_0$, as fixed (by conditioning on their values). Define additional variables $V_i \stackrel{\text{iid}}{\sim} \mathsf{Unif}[0,1]$, independent from the $p_i$'s. Define also

$$B_i = \mathbb{1}\{V_i \le h(p_i)/C\} \ .$$

Write $p_{1:n}$ to denote $p_1, \ldots, p_n$. Note that, conditioning on $p_{1:n}$, we then have that the $B_i$'s are independent, with distributions

$$(B_i \mid p_{1:n}) \stackrel{\perp\!\!\!\perp}{\sim} \mathsf{Bernoulli}\left(\frac{h(p_i) \wedge C}{C}\right) \ . \tag{19}$$

Furthermore, marginally, we see that for all $i \in \mathcal{H}_0$, the $B_i$'s are independent Bernoulli variables with

$$\mathbb{E}[B_i] = \mathbb{E}\left[\frac{h(p_i) \wedge C}{C}\right] = \frac{1}{C}\int_{t=0}^{1} \mathsf{h}(p_i) \wedge C \ \mathsf{d}t =: \rho \ .$$

Next, we would like to apply Lemma 4 to bound $\mathbb{E}\left[\frac{1+\#\{i \le \widehat{k} : i \in \mathcal{H}_0\}}{1+\sum_{i \le \widehat{k}, i \in \mathcal{H}_0} B_i}\right]$. First, we need to construct a filtration $\{\mathcal{F}_k\}_{k=1,\ldots,n}$ that satisfies the conditions of this lemma.

Let $\mathcal{F}_k$ be the $\sigma$-algebra defined by knowing $(p_i, V_i)$ for all $i \notin \mathcal{H}_0$, knowing $(p_i, V_i)$ for all $i > k$ with $i \in \mathcal{H}_0$, and knowing $\{(p_i, V_i) : i \le k, i \in \mathcal{H}_0\}$ (note that this is an unordered set—for instance, if $1 \in \mathcal{H}_0$ and $2 \in \mathcal{H}_0$, then $\mathcal{F}_2$ knows $(p_1, V_1)$ and $(p_2, V_2)$ but does not know which one is which). For each $i$, $B_i$ is a function of $(p_i, V_i)$. Furthermore, the $(p_i, V_i)$'s are i.i.d. for $i \in \mathcal{H}_0$. Therefore, this filtration satisfies the conditions (16), (17), and (18) of Lemma 4. Furthermore, $\widehat{k}$ as defined in (15) is a stopping time with respect to $\{\mathcal{F}_t\}$. Combining Lemma 4 with the Optional Stopping Theorem, therefore,

$$\mathbb{E}\left[\frac{1 + \#\{i \le \widehat{k} : i \in \mathcal{H}_0\}}{1 + \sum_{i \le \widehat{k}, i \in \mathcal{H}_0} B_i}\right] \le \frac{1}{\rho} = \frac{1}{\frac{1}{C}\int_{t=0}^{1} \mathsf{h}(p_i) \wedge C \ \mathsf{d}t} \ . \tag{20}$$

Next, we calculate

$$\begin{aligned}
&\mathbb{E}\left[\frac{1+\#\{i\leq\widehat{k}:i\in\mathcal{H}_0\}}{1+\sum_{i\leq\widehat{k},i\in\mathcal{H}_0}B_i}\right]\\
&=\mathbb{E}\left[\mathbb{E}\left[\frac{1+\#\{i\leq\widehat{k}:i\in\mathcal{H}_0\}}{1+\sum_{i\leq\widehat{k},i\in\mathcal{H}_0}B_i}\;\middle|\;p_{1:n}\right]\right] \quad \text{by the tower rule of expectations}\\
&=\mathbb{E}\left[(1+\#\{i\leq\widehat{k}:i\in\mathcal{H}_0\})\cdot\mathbb{E}\left[\frac{1}{1+\sum_{i\leq\widehat{k},i\in\mathcal{H}_0}B_i}\;\middle|\;p_{1:n}\right]\right] \quad \text{since } \widehat{k} \text{ is a function of } p_{1:n}\\
&\geq\mathbb{E}\left[(1+\#\{i\leq\widehat{k}:i\in\mathcal{H}_0\})\cdot\frac{1}{\mathbb{E}\left[1+\sum_{i\leq\widehat{k},i\in\mathcal{H}_0}B_i\;\middle|\;p_{1:n}\right]}\right] \quad \text{by Jensen's inequality}\\
&=\mathbb{E}\left[(1+\#\{i\leq\widehat{k}:i\in\mathcal{H}_0\})\cdot\frac{1}{1+\sum_{i\leq\widehat{k},i\in\mathcal{H}_0}\frac{h(p_i)\wedge C}{C}}\right] \quad \text{by (19)}\\
&\geq C\cdot\mathbb{E}\left[(1+\#\{i\leq\widehat{k}:i\in\mathcal{H}_0\})\cdot\frac{1}{C+\sum_{i\leq\widehat{k},i\in\mathcal{H}_0}h(p_i)}\right].
\end{aligned}$$

Combining this result with (20), we have proved the lemma. □

## 4.2 Proof of Theorem 2 (asymptotic FDR control)

*Proof of Theorem 2.* Take any sequence $C_n > 0$ with $C_n \to \infty$ and $\frac{C_n}{m_n} \to 0$ (for instance, we may take $C_n = \sqrt{m_n}$). Then note that

$$\lim_{n\to\infty}\int_{t=0}^{1}\mathsf{h}(t)\wedge C_n\ \mathsf{d}t = 1$$

because we know that $\int_{t=0}^{1}\mathsf{h}(t)\ dt = 1$ by assumption. We then have

$$\begin{aligned}
\mathbb{E}\left[\mathsf{FDP}(\widehat{k}_{\mathsf{h}})\right] &= \mathbb{E}\left[\mathsf{FDP}(\widehat{k}_{\mathsf{h}})\cdot\mathbb{1}\left\{\widehat{k}_{\mathsf{h}}<m_n\right\}\right]+\mathbb{E}\left[\mathsf{FDP}(\widehat{k}_{\mathsf{h}})\cdot\mathbb{1}\left\{\widehat{k}_{\mathsf{h}}\geq m_n\right\}\right]\\
&\leq \mathbb{P}\left\{\widehat{k}_{\mathsf{h}}<m_n\right\}+\mathbb{E}\left[\frac{\mathsf{FalsePos}(\widehat{k}_{\mathsf{h}})}{\widehat{k}_{\mathsf{h}}}\cdot\mathbb{1}\left\{\widehat{k}_{\mathsf{h}}\geq m_n\right\}\right]\\
&\leq \mathbb{P}\left\{\widehat{k}_{\mathsf{h}}<m_n\right\}+\mathbb{E}\left[\frac{\mathsf{FalsePos}(\widehat{k}_{\mathsf{h}})}{C_n/\alpha+\widehat{k}_{\mathsf{h}}}\right]\cdot\frac{C_n/\alpha+m_n}{m_n}\\
&= \mathbb{P}\left\{\widehat{k}_{\mathsf{h}}<m_n\right\}+\mathbb{E}\left[\mathsf{mFDP}_{C_n/\alpha}(\widehat{k}_{\mathsf{h}})\right]\cdot\frac{C_n/\alpha+m_n}{m_n}\\
&\leq \mathbb{P}\left\{\widehat{k}_{\mathsf{h}}<m_n\right\}+\frac{\alpha}{\int_{t=0}^{1}\mathsf{h}(t)\wedge C_n\ \mathsf{d}t}\cdot\frac{C_n/\alpha+m_n}{m_n} \quad \text{by Theorem 1}.
\end{aligned}$$

Taking limits,

$$\lim_{n\to\infty} \mathbb{E}\left[\mathsf{FDP}(\widehat{k}_{\mathsf{h}})\right] \le \lim_{n\to\infty} \left(\mathbb{P}\left\{\widehat{k}_{\mathsf{h}} < m_n\right\} + \frac{\alpha}{\int_{t=0}^1 \mathsf{h}(t)\wedge C_n \,\mathsf{d}t} \cdot \frac{C_n/\alpha + m_n}{m_n}\right)$$
$$= \lim_{n\to\infty} \mathbb{P}\left\{\widehat{k}_{\mathsf{h}} < m_n\right\} + \frac{\alpha}{\lim_{n\to\infty}\int_{t=0}^1 \mathsf{h}(t)\wedge C_n \,\mathsf{d}t} \cdot \lim_{n\to\infty} \frac{C_n/\alpha + m_n}{m_n}$$
$$= 0 + \frac{\alpha}{1}\cdot 1 = \alpha\ .$$

□

## 4.3 Proof of Theorem 3 (asymptotic power calculation)

We begin by stating a preliminary lemma on maximal values for random walks:

**Lemma 5.** *Let $X_1, X_2, \dots$ be independent random variables with $\mathbb{E}[X_i] = 0$, and such that $X_i$ is $(\sigma^2, b)$-subexponential (as defined in (11)) for all $i$. Then for any $\epsilon \in (0,1)$,*

$$\mathbb{P}\left\{\left|\sum_{i=1}^t X_i\right| \le \sqrt{2\log_2(4/\epsilon)}\cdot \max\left\{\sigma, b\sqrt{2\log_2(4/\epsilon)}\right\}\cdot \sqrt{t\log(1+t)} \text{ for all } t\ge 1\right\} \ge 1-\epsilon\ .$$

This lemma is proved in Appendix A.2. With this result in place, we turn to the proof of the theorem.

*Proof of Theorem 3.* First, we define the asymptotic expected FDP estimate along the sequence of p-values,

$$\mathsf{E}(t) = 1 - f(t)\cdot(1-\mu)\ .$$

That is, at the cutoff $k = t\cdot n$, we expect that $\widehat{\mathsf{FDP}}_{\mathsf{h}}(k) \approx \mathsf{E}(t)$. Note that $\mathsf{E}(t)$ is monotone nondecreasing, due to the assumption that $f(t)$ is nonincreasing.

Next, we prove that the approximation $\widehat{\mathsf{FDP}}_{\mathsf{h}}(k) \approx \mathsf{E}(t)$ is uniformly accurate. Fix any $k \in \{1,\dots,n\}$. First we consider the expectation:

$$\mathbb{E}\left[\frac{\sum_{i=1}^k h(p_i)}{k}\right] = \mathbb{E}\left[\frac{k - \sum_{i=1}^k (1-h(p_i))}{k}\right] = 1 - \frac{\#\{i\le k : i\notin \mathcal{H}_0\}}{k}\cdot(1-\mu)\ ,$$

and so, applying (12),

$$\left|\mathbb{E}\left[\frac{\sum_{i=1}^k h(p_i)}{k}\right] - \mathsf{E}\left(\frac{k}{n}\right)\right| = \left|\frac{\#\{i\le k : i\notin\mathcal{H}_0\}}{k} - f\left(\frac{k}{n}\right)\right|\cdot(1-\mu) \le (1-\mu)\epsilon_n\ . \quad (21)$$

Next, we apply Lemma 5 to prove that $\widehat{\mathsf{FDP}}_{\mathsf{h}}(k) = \frac{\sum_{i=1}^k h(p_i)}{k} \approx \mathbb{E}\left[\frac{\sum_{i=1}^k h(p_i)}{k}\right]$ for all sufficiently large $k$. Applying Lemma 5 with $\epsilon = \frac{1}{\log(n)}$, we see that with probability at least $1 - \frac{1}{\log(n)}$,

$$\left|\frac{\sum_{i=1}^k h(p_i)}{k} - \frac{\sum_{i=1}^k \mathbb{E}[h(p_i)]}{k}\right| \le$$
$$\sqrt{2\log_2(4\log(n))}\cdot\max\left\{\sigma, b\sqrt{2\log_2(4\log(n))}\right\}\cdot\sqrt{\frac{\log(1+k)}{k}} \text{ for all } k\ge 1\ , \quad (22)$$

and therefore, restricting our attention to $k > \log(n)$ and combining with our result in (21), we have

$$\max_{k>\log(n)} \left| \widehat{\mathsf{FDP}}_{\mathsf{h}}(k) - \mathsf{E}\left(\frac{k}{n}\right) \right| <$$
$$\sqrt{2\log_2(4\log(n))} \cdot \max\left\{\sigma, b\sqrt{2\log_2(4\log(n))}\right\} \cdot \sqrt{\frac{\log(1+\log(n))}{\log(n)}} + (1-\mu)\epsilon_n =: \beta_n \ . \tag{23}$$

Note that $\beta_n \to 0$. We also define $\tau_n = \frac{\beta_n}{(1-\mu)\delta}$, where the constant $\delta$ is from assumption (13), and note that $\tau_n \to 0$ also.

Now we split into cases. Here we treat the case that $f(1) < \frac{1-\alpha}{1-\mu} < f(0)$, and defer the other two cases to the Appendix.

**Case 1: $\alpha$ satisfies $f(1) < \frac{1-\alpha}{1-\mu} < f(0)$.** In this case, we will prove that, if (22) holds, then

$$\begin{cases} \widehat{\mathsf{FDP}}_{\mathsf{h}}(k) \le \alpha \text{ for all } \log(n) < k \le n\cdot(T-\tau_n), \text{ and} \\ \widehat{\mathsf{FDP}}_{\mathsf{h}}(k) > \alpha \text{ for all } k > \max\{\log(n), n\cdot(T+\tau_n)\} \ . \end{cases} \tag{24}$$

If this holds, then by definition of $\widehat{k}_{\mathsf{h}}$, this implies that for sufficiently large $n$,

$$n\cdot(T-\tau_n) \le \widehat{k}_{\mathsf{h}} \le n\cdot(T+\tau_n) \ ,$$

and therefore,

$$\frac{\#\{i \le n\cdot(T-\tau_n) : i \notin \mathcal{H}_0\}}{N_1(n)} \le \mathsf{Power} \le \frac{\#\{i \le n\cdot(T+\tau_n) : i \notin \mathcal{H}_0\}}{N_1(n)} \ .$$

Using assumption (12), therefore,

$$\frac{n\cdot(T-\tau_n)\cdot(f(T-\tau_n)-\epsilon_n)}{n\cdot(f(1)+\epsilon_n)} \le \mathsf{Power} \le \frac{n\cdot(T+\tau_n)\cdot(f(T+\tau_n)+\epsilon_n)}{n\cdot(f(1)-\epsilon_n)} \ . \tag{25}$$

Since the limit of both sides is equal to $T\cdot\frac{f(T)}{f(1)}$, this proves the desired result. To be more precise, we have proved that the bound (25) holds with probability at least $1-\frac{1}{\log(n)}$ (since this is a lower bound on the probability of the event (22)), which itself tends to 1. Therefore, the power of our procedure converges to the limit $T\cdot\frac{f(T)}{f(1)}$ in probability.

It remains to be shown that (22) implies (24). First, note that $f(T) = \frac{1-\alpha}{1-\mu} > 1-\alpha$, and so, since $\tau_n \to 0$ and $f$ is continuous, we see that

$$\min_{t\in[T-\tau_n, T+\tau_n]} f(t) \ge 1-\alpha$$

for sufficiently large $n$. Therefore, by assumption (13),

$$f'(t) \le -\delta \text{ for all } t \in [T-\tau_n, T+\tau_n] \ .$$

Now take any $k$ such that $\log(n) < k \le n\cdot(T-\tau_n)$. Then

$$\widehat{\mathsf{FDP}}_{\mathsf{h}}(k) \le \mathsf{E}\left(\frac{k}{n}\right) + \beta_n \le \mathsf{E}(T-\tau_n) + \beta_n = 1 - f(T-\tau_n)\cdot(1-\mu) + \beta_n$$
$$\le 1 - (f(T) + \tau_n\cdot\delta)\cdot(1-\mu) + \beta_n = 1 - f(T)\cdot(1-\mu) = \alpha \ ,$$

where the first inequality applies (23). This proves the first part of (24); the second part of (24) is proved similarly. The remaining cases are proved in Appendix A.4.

□

# 5 Simulations

In this section we evaluate the performance of various accumulation tests on two tasks with simulated data: a ranked hypothesis testing problem (Section 5.1), and a high-dimensional linear regression problem (Section 5.2). Code to reproduce the first simulated data experiment is available online.[2]

## 5.1 Simulations for the sequential testing problem

Here we examine our accumulation test under four different simulation settings, and compare the performance of several accumulation test methods: SeqStep with parameter $C = 2$, SeqStep+ with $C = 2$, ForwardStop, and the new HingeExp method with parameter $C = 2$ (see Section 2 for the definitions of these methods). The sequences of hypotheses and of p-values in our simulations vary on the degree of separation between the nulls and the non-nulls (the extent to which non-null hypotheses concentrate early in the list), and on the signal strength of the non-nulls (the extent to which non-null p-values are visibly different from a uniform distribution).

### 5.1.1 Methods

To create the simulated data, we generate the sequence of p-values for $n = 1000$ hypotheses with 100 non-nulls, by the following steps:

1. First, we generate "prior information" for each hypothesis. We draw z-scores $Z_i$ independently, with $Z_i$ drawn from $N(0, 1)$ for nulls $i \in \mathcal{H}_0$ and from $N(\mu_1, 1)$ for non-nulls $i \notin \mathcal{H}_0$. Here $\mu_1 > 0$ controls the extent of the separation between nulls and non-nulls.

2. Sort the z-scores in descending order according to magnitude: $|Z_{(1)}| > |Z_{(2)}| > \cdots > |Z_{(n)}|$. Assign a new index to each hypothesis, according to its position in the sorted list.

3. Now we generate p-values for each hypothesis. We draw new z-scores $Z_i^*$ independently for each hypothesis, with $Z_i^* \sim N(0, 1)$ for nulls $i \in \mathcal{H}_0$ and $Z_i^* \sim N(\mu_2, 1)$ for non-nulls $i \notin \mathcal{H}_0$. Here $\mu_2 > 0$ controls the strength of the true signals. Then we calculate p-values with a two-tailed z-test, $p_i = 2(1 - \Phi(|Z_i^*|))$, where $\Phi$ is the cumulative distribution function of the standard normal distribution. Note that these p-values are independent from the process of ranking the hypotheses.

The ranking (and separation) of nulls and non-nulls is determined in steps 1 and 2, and controlled by $\mu_1$. Larger $\mu_1$ leads to better separation between nulls and non-nulls. The strength of non-null signals is specified in step 3, and controlled by $\mu_2$. As $\mu_2$ gets larger, the signals become stronger. For settings with good separation of nulls and non-nulls and with strong signals, it is easier to achieve high power while keeping FDR controlled. Here, we choose four settings, with two levels of separation $\mu_1 \in \{2, 3\}$, and two levels of signal strength $\mu_2 \in \{2, 3\}$.

[2] http://www.stat.uchicago.edu/~rina/accumulationtests.html

Under each simulation setting, we compare the performance of the four selected accumulation test methods. In each trial, the power and FDP for each accumulation function and rejection rule are recorded, over a range of target FDR levels $\alpha \in \{0.05, 0.075, \ldots, 0.25\}$. We also record the estimated FDP in each trial as we move along the list of hypotheses, $\widehat{\mathsf{FDP}}_{\mathsf{h}}(k)$ , for each of the four methods, and compare to the true false discovery proportion $\mathsf{FDP}(k)$, as $k$ ranges from 1 to $n$. The performance results are averaged over 50 trials.

### 5.1.2 Results

Figure 2 shows the average power and average observed FDR of the four selected accumulation test methods, plotted against the target FDR level $\alpha$. In the weak signal regime ($\mu_2 = 2$), the methods are mostly conservative in terms of FDR, with an observed FDR that is lower than the target level $\alpha$, and thus power is low—this is expected, since the non-null p-values contribute to the estimated false discovery proportion. In contrast, with strong signal ($\mu_2 = 3$), the observed FDR levels are closer to $\alpha$, and in fact HingeExp has slightly higher FDR than desired when separation is poor ($\mu_1 = 2$)—this is not unexpected, since HingeExp only guarantees the control of modified FDR (see Lemma 1). The power of the methods improves with stronger signal (larger $\mu_2$) and with better separation (larger $\mu_1$). Across all settings, HingeExp consistently gives the highest average power and observed FDR, while SeqStep+ is generally the most conservative, with lowest average power and observed FDR.

To further explore the differences between the methods, we also compare the estimated false discovery proportion along the list of hypotheses, $\widehat{\mathsf{FDP}}_{\mathsf{h}}(k)$ for $k = 1, \ldots, n$, for each of the four methods, and compare with the actual false discovery proportion, $\mathsf{FDP}(k)$. (For the SeqStep+ method we define $\widehat{\mathsf{FDP}}_{\mathsf{h}}(k)$ to agree with the method definition (4).) Figure 3 shows the results, averaged over 50 simulations. For settings with stronger signals (i.e. $\mu_2 = 3$), $\widehat{\mathsf{FDP}}_{\mathsf{h}}(k)$ is a good estimate of $\mathsf{FDP}(k)$, while for settings with weak signals (e.g. $\mu_2 = 2$), $\widehat{\mathsf{FDP}}_{\mathsf{h}}(k)$ overestimates $\mathsf{FDP}(k)$, as expected. Comparing across methods, the HingeExp method function yields the estimate $\widehat{\mathsf{FDP}}_{\mathsf{h}}(k)$ that approximates the actual FDP best.

## 5.2 Simulations for the least angle regression (LARS) path

Inference for high dimensional regression has been a problem of wide interest in many modern applications. Recently, Taylor et al. [37] proposed the spacing test method for post-selection inference of LARS (closely related to the commonly used LASSO method for penalized sparse regression [38]), which gives a p-value for each feature in the order of being selected in the LARS path. The $i$th p-value, $p_i$, is distributed as $\mathsf{Unif}[0, 1]$, under the null hypothesis that the partial regression coefficient of the $i$th variable in the LARS path is 0, in the model containing all active variables at the $i$th LARS step; furthemore, the p-values are independent. This provides an ordered list of p-values that follows the assumptions of ordered hypothesis testing problem, and therefore, can be treated with the accumulation method. (Note, however, that the null hypothesis refers to the *partial* regression coefficient, while we may often be interested in the coefficient in the true model.) As the sequence corresponds to signal and noisy features in the LARS path, the test provides a stopping rule for LARS with guarantee on FDR level (here $\mathsf{FDP}(k)$ is the proportion of noise among all features included up to the $k$th LARS step). In this simulation, compare the performance of the SeqStep (with parameter $C = 2$), SeqStep+ (with $C = 2$), ForwardStop, and HingeExp (with $C = 2$) accumulation test methods, under three settings of feature signal strength.

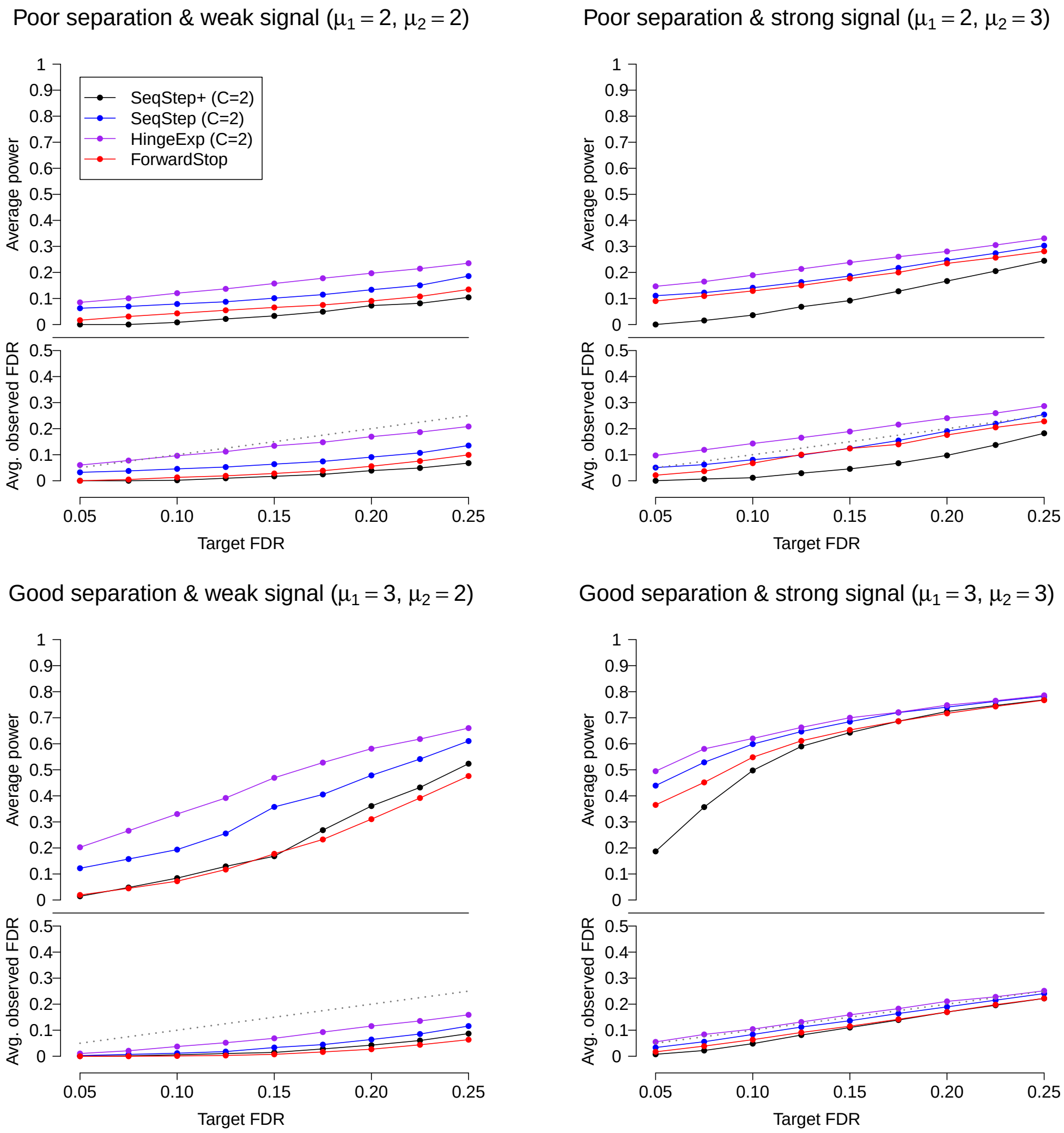

*Figure 2: Power and observed FDR level of the SeqStep, SeqStep+, ForwardStop, and HingeExp methods, plotted against target FDR level $\alpha$ (averaged over 50 trials).*

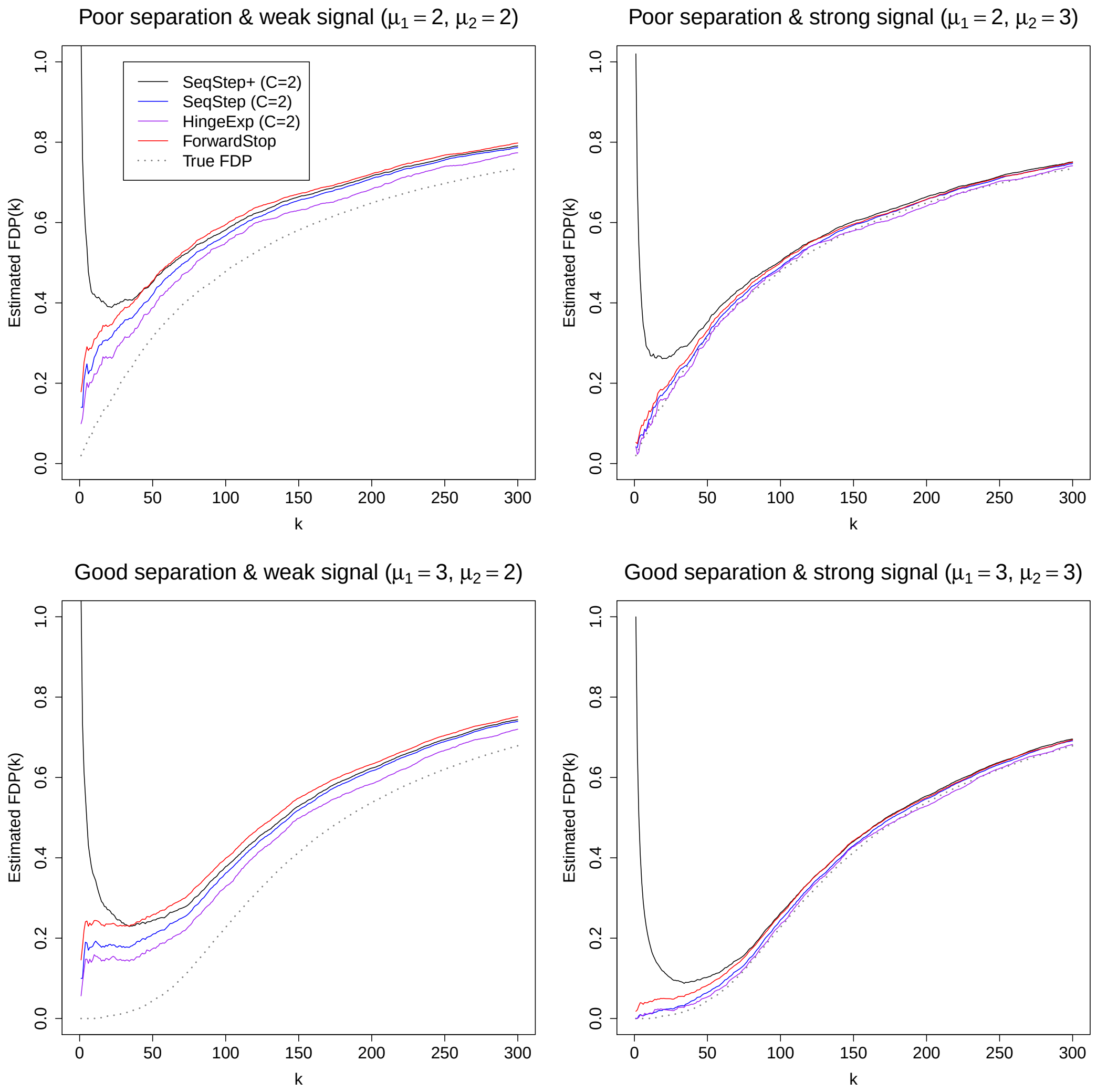

*Figure 3: Estimated FDP with the SeqStep, SeqStep+, ForwardStop, and HingeExp methods, plotted against the true FDP, across $k = 1, \ldots, p$ (results are averaged over 50 trials).*

### 5.2.1 Methods

In all three simulation settings, there are $N = 200$ observations on $p = 100$ features, of which either $k^\star = 10$ or $k^\star = 20$ are true signals (nonzero coefficients) The design matrix $X$ consists of i.i.d. standard normal entries. The nonzero signals $\beta_j$, for features $j = 1, \dots, k^\star$, are taken to be equally spaced value from $2 \cdot \gamma$ to $\sqrt{2\log(p)} \cdot \gamma$, where $\gamma = 1, 3, 5$ in the three settings. This forms a gradient from weak signal to strong signal scenarios. The remaining coefficients are set as $\beta_{k^\star+1} = \cdots = \beta_p = 0$. The response is then generated as

$$y = X\beta + \epsilon \ ,$$

where the entries of $\epsilon$ are also i.i.d. standard normal variables. Given the simulated data $X, y$, the LARS method and the spacing test are applied, yielding p-values for sequential testing. Note that the number of hypotheses is now given by $p$ (one for each feature), rather than the former notation $n$.

### 5.2.2 Results

Figure 4 shows the average power and observed FDR of the four accumulation tests, averaged over 50 trials. When $\gamma = 1, 3$, all four methods successfully control FDR, and when $\gamma = 5$, SeqStep+ and ForwardStop control FDR well, while HingeExp and SeqStep slightly exceed the target FDR level $\alpha$. In all settings, HingeExp attains the highest average power and observed FDR level, while SeqStep+ is extremely conservative for lower $\alpha$ values (due to the fact that, with only $k^\star = 10$ or $k^\star = 20$ true signals, few discoveries are made overall, so the slightly conservative correction in this method (4) has a large effect).

We also plot the estimated false discovery proportion, $\widehat{\mathsf{FDP}}_{\mathsf{h}}(k)$ over the first $k$ steps of the LARS path ($k = 1, \dots, p$), against the actual $\mathsf{FDP}(k)$. Figure 5 shows the results, averaged over 50 simulations. As expected, the estimated FDP levels increasingly overestimate the true FDP as signal strength $\gamma$ decreases. SeqStep+ is quite conservative due to the correction term in the method's definition, while the other three methods show no consistent trend in terms of accurate estimation of $\mathsf{FDP}(k)$.

## 6 Application to dosage response data

We now show an application of our methods to the problem of identifying effects of drug dosage on gene expression levels. Code to reproduce this real data experiment is available online.[3]

Suppose that we measure the following data: gene expression levels for genes $i = 1, \dots, n$ are measured in $m = m_{\mathsf{C}} + m_{\mathsf{L}} + m_{\mathsf{H}}$ independent trials, where the trials $\{1, \dots, m\}$ are partitioned into three sets $T_{\mathsf{C}} = \{1, \dots, m_{\mathsf{C}}\}$, $T_{\mathsf{L}} = \{m_{\mathsf{C}}+1, \dots m_{\mathsf{C}}+m_{\mathsf{L}}\}$, and $T_{\mathsf{H}} = \{m_{\mathsf{C}}+m_{\mathsf{L}}+1, \dots, m_{\mathsf{C}}+m_{\mathsf{L}}+m_{\mathsf{H}}\}$, such that:

- For each $j \in T_{\mathsf{C}}$, trial $j$ is carried out in the absence of the drug (the control group);
- For each $j \in T_{\mathsf{L}}$, trial $j$ is carried out under a low drug dosage; and
- For each $j \in T_{\mathsf{H}}$, trial $j$ is carried out under a high drug dosage.

[3] http://www.stat.uchicago.edu/~rina/accumulationtests.html

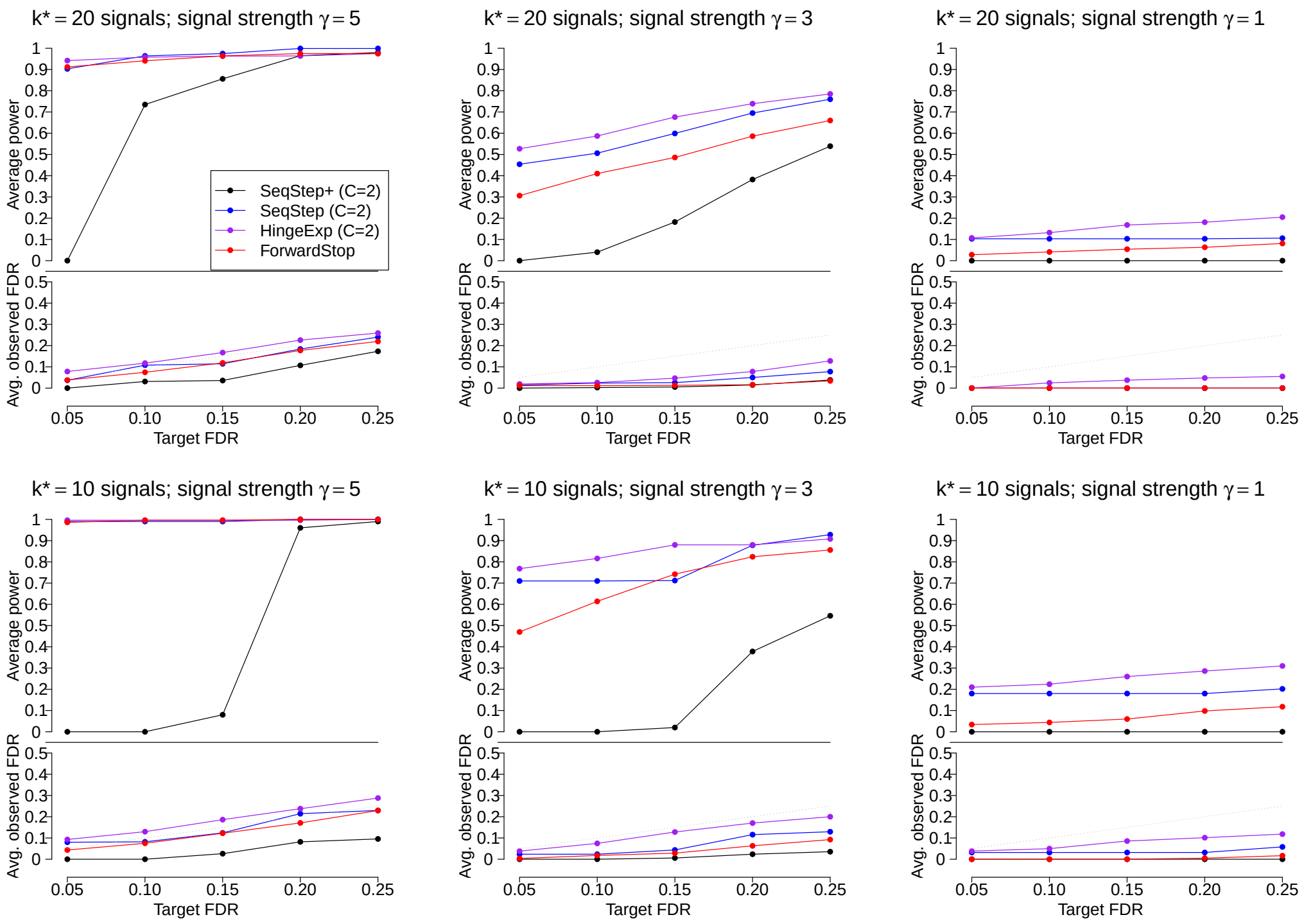

*Figure 4: Power and observed FDR level of the SeqStep, SeqStep+, ForwardStop, and HingeExp methods for the LARS path, plotted against target FDR level $\alpha$ (averaged over 50 trials).*

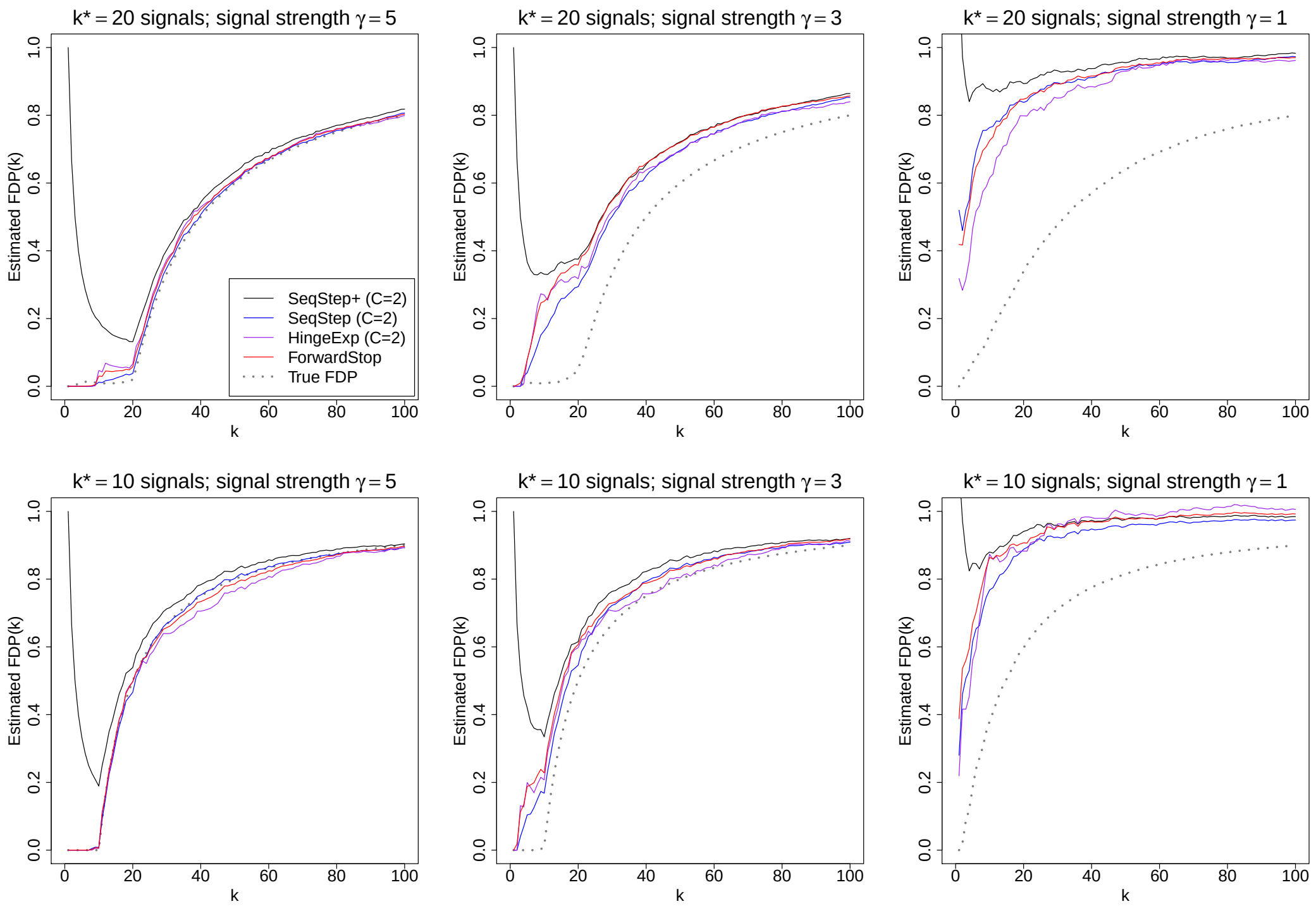

*Figure 5: Estimated FDP with the SeqStep, SeqStep+, ForwardStop, and HingeExp methods for the LARS path, plotted against the true FDP, across $k = 1, \ldots, p$ (results are averaged over 50 trials).*

For each gene $i$ and each trial $j$, let $X_{ij} \in \mathbb{R}$ be the logarithm of the measured gene expression level of gene $i$ in trial $j$.

In this type of setting, we may potentially be interested in various questions related to estimating the effect of dosage on gene expression level and testing for the significance of these effects. In general, identifying genes that respond to the higher dosage will be easier than at the lower dosage, since the magnitude of the response will often depend on the dosage used. To be specific, for each gene $i$ we are interested in testing the null hypothesis $H_i$, which states that the observations

$$\{X_{ij} : j \in T_{\mathsf{C}} \cup T_{\mathsf{L}}\}$$

are i.i.d. (in other words, the low dose has no effect on the distribution for the gene expression level for gene $i$). For simplicity in the discussion and analysis below, we treat the measurements for each gene as though they were independent—of course, this does not hold in practice, and in future work we hope to develop results on FDR control of the accumulation tests under p-value dependence. For the present experiment, each of the methods we compare here comes with theoretical guarantees of FDR control only under the independence assumption.[4]

In order to identify which genes are differentially expressed at the low dosage level (as compared to the control), we can take several different approaches:

- Two-sample approach. First, we can take the simple approach of comparing only the control data and the low dosage data, while discarding the high dosage data. For each gene $i$, we would calculate a p-value $p_i$ by comparing the control group observations $\{X_{ij} : j \in T_{\mathsf{C}}\}$ with the low dosage observations $\{X_{ij} : j \in T_{\mathsf{L}}\}$. To control FDR, we could then apply the Benjamini-Hochberg procedure [3]. This approach has the drawback that it makes no use of the valuable information in the high dosage trials; in our experiment, we show that this approach has relatively low power.
- Joint model approach. At the other extreme, we could construct a joint model for gene responses at each of the dosage levels, and fit this model to the entire data set. Using the entire data set, including the high dosage data, would increase our power to detect differential responses at the low dosage since information is shared across these two conditions. However, the validity of the inference we perform could be extremely sensitive to the choice of the model.
- Sequential testing approach. Finally, we may apply a sequential testing procedure, which combines the benefits of the two approaches above. In our experiment, we will see that this approach will control FDR, while gaining substantial power as compared to the first approach where the high dosage data is discarded.

For the remainder of this section, for two sets of observations $A$ and $B$, define $\mathsf{Pval}(A, B)$ to be the p-value produced by a two-sided two-sample t-test comparing the observations in $A$ with the observations in $B$ (assuming unequal variances between the two populations). Define $\mathsf{Pval}_+(A, B)$ and $\mathsf{Pval}_-(A, B)$ analogously for one-sided two-sample t-tests, where $\mathsf{Pval}_+(A, B)$ tests for evidence that the mean of $A$'s population is larger than the mean of $B$'s population, and $\mathsf{Pval}_-(A, B)$ does the reverse.

[4] While several existing methods for multiple testing yield guarantees for FDR control even in the case of dependent p-values, the methods we are aware of are quite conservative and yield nearly zero power in this experiment. Specifically, Benjamini and Yekutieli's modification [5] of the BH method, the Holm-Bonferroni method [25], and the Bonferroni correction each yielded, even at target FDR level $\alpha = 0.9$, no more than two discoveries for this gene expression experiment; in contrast, the accumulation test methods examined here yield thousands of discoveries.

We now follow these steps to reformulate the dosage/gene expression problem as a sequential hypothesis testing problem:

1. For each gene $i$, calculate $p_i^{\mathsf{high}}$ as

$$p_i^{\mathsf{high}} = \mathsf{Pval}\left(\{X_{ij} : j \in T_{\mathsf{H}}\}, \{X_{ij} : j \in T_{\mathsf{C}} \cup T_{\mathsf{L}}\}\right) \ .$$

   Record also $s_i \in \{+, -\}$, the sign of the estimated effect, i.e.

$$s_i = \mathrm{sign}\left(\frac{1}{m_{\mathsf{H}}}\sum_{j \in T_{\mathsf{H}}} X_{ij} - \frac{1}{m_{\mathsf{C}} + m_{\mathsf{L}}}\sum_{j \in T_{\mathsf{C}} \cup T_{\mathsf{L}}} X_{ij}\right) \ .$$

2. We then use these high-dosage p-values to relabel the $n$ genes, that is, reorder the genes so that

$$p_1^{\mathsf{high}} \le p_2^{\mathsf{high}} \le \dots \le p_n^{\mathsf{high}} \ .$$

3. Next, for each gene $i$, compute an initial p-value by comparing the low-dosage trials with the control trials. We use a one-sided two-sample t-test (determined by the sign $s_i$):

$$p_i^{\mathsf{init}} = \mathsf{Pval}_{s_i}\left(\{X_{ij} : j \in T_{\mathsf{L}}\}, \{X_{ij} : j \in T_{\mathsf{C}}\}\right) \ .$$

   The reason that we use a one-sided t-test (rather than two-sided), is that, if for instance we observe a positive response at the high dosage for gene $i$, then we are much more likely to see a positive (rather than negative) effect at the low dosage, as well. Therefore, using a one-sided t-test is likely to achieve higher power than a two-sided test.

4. Now we transform to the final p-values using a permutation test. For each gene $i$, for each possible permutation $\pi$ on the trial labels $\{1, \dots, m_{\mathsf{C}} + m_{\mathsf{L}}\}$, compute

$$p_i^{\pi} = \mathsf{Pval}_{s_i}\left(\{X_{i\pi(j)} : j \in T_{\mathsf{L}}\}, \{X_{i\pi(j)} : j \in T_{\mathsf{C}}\}\right) \ .$$

   We then calculate the final p-value by comparing $p_i^{\mathsf{init}}$ with the empirical distribution $\{p_i^{\pi}$ : all possible permutations $\pi\}$:

$$p_i = \frac{\#\{\pi \ : \ p_i^{\mathsf{init}} \le p_i^{\pi}\}}{(m_{\mathsf{C}} + m_{\mathsf{L}})!} \ , \tag{26}$$

   and perform the accumulation test on this sequence of p-values.

In fact, since $p_i^{\pi}$ depends only on the partition of the $m_{\mathsf{C}} + m_{\mathsf{L}}$ many trial labels into two groups of size $m_{\mathsf{C}}$ and $m_{\mathsf{L}}$ (the control group and the low-dose group), we only need to calculate $p_i^{\pi}$ for $\binom{m_{\mathsf{C}}+m_{\mathsf{L}}}{m_{\mathsf{C}}}$ many permutations.

Note that $p_i^{\mathsf{high}}$ and $s_i$ depend on $\{X_{ij} : j \in T_{\mathsf{C}} \cup T_{\mathsf{L}}\}$, and so the reordering is not independent of this data. Therefore, we cannot use the t-test p-values $p_i^{\mathsf{init}}$ for the accumulation test. However, $p_i^{\mathsf{high}}$ and $s_i$ are invariant to permutations of this input by definition of the two-sample t-test. In other words, even after conditioning on $(p_i^{\mathsf{high}}, s_i)$, the variables $\{X_{ij} : j \in T_{\mathsf{C}} \cup T_{\mathsf{L}}\}$ are exchangeable under the null hypothesis $H_i$. Therefore, even after reordering the genes according to the high-dosage p-values $p_i^{\mathsf{high}}$ and recording signs $s_i$, the final permutation test p-values $p_i$ that we calculate are valid p-values for each true null hypothesis $H_i$. Our theory thus guarantees FDR control when the accumulation test is applied to these permutation test p-values (if we assume that the data for each gene is independent).

## 6.1 Empirical results

We now implement the methods described above on real data. All computations are carried out in R [29].

The data[5] [10] measures differential expression in response to estrogen in breast cancer cells. The data set consists of $n = 22283$ genes with 25 trials, with 5 trials each for the control group and for four different dosage levels. For our experiment, we use the $m_{\mathsf{C}} = 5$ control trials, the $m_{\mathsf{L}} = 5$ trials at the lowest dosage, and the $m_{\mathsf{H}} = 5$ trials at the highest dosage.

We compare the following methods, each with target FDR level $\alpha = 0, 0.01, 0.02, \ldots, 0.90$:

- The accumulation test applied to the sequence of one-sided t-test p-values $p_i$ (26), using one of the following accumulation functions:

$$\mathsf{h}(t) = 2 \cdot \mathbb{1}\{t > 0.5\}\ ,\ \mathsf{h}(t) = 2 \cdot \log\left(\frac{1}{2(1-t)}\right) \cdot \mathbb{1}\{t > 0.5\}\ ,\ \mathsf{h}(t) = \log\left(\frac{1}{1-t}\right)\ ,$$

  corresponding to the SeqStep method with $C = 2$, the HingeExp function with $C = 2$, and the ForwardStop method[6], respectively. We also test SeqStep+ (recall (4)) with $C = 2$.

- The Benjamini-Hochberg procedure [3], using p-values that compare the low dosage trials with the control trials, via either a two-sided t-test,

$$p_i^{\mathsf{ttest}} = \mathsf{Pval}\left(\{X_{ij} : j \in T_{\mathsf{L}}\}, \{X_{ij} : j \in T_{\mathsf{C}}\}\right)\ , \tag{27}$$

  or a the permutation test on these t-tests,

$$p_i^{\mathsf{perm}} = \frac{\#\{\pi\ :\ p_i^{\mathsf{ttest}} \leq p_i^{\pi}\}}{(m_{\mathsf{C}} + m_{\mathsf{L}})!} \text{ where } p_i^{\pi} = \mathsf{Pval}\left(\{X_{i\pi(j)} : j \in T_{\mathsf{L}}\}, \{X_{i\pi(j)} : j \in T_{\mathsf{C}}\}\right)\ . \tag{28}$$

- Storey [34]'s modification of the Benjamini-Hochberg procedure, applied to either the t-test p-values (27) or the permutation test p-values (28). Since the value of $\alpha$ ranges from 0 to 0.9, we estimate the number of true nulls as $\widehat{m}_0 = 10 \cdot \#\{i : p_i^{\mathsf{ttest}} > 0.9\}$ or $\widehat{m}_0 = 10 \cdot \#\{i : p_i^{\mathsf{perm}} > 0.9\}$, respectively, for the two types of p-values.

For the various methods, Figure 6 displays the number of discoveries against the target FDR level $\alpha$. We see that the accumulation tests far outperform the Benjamini-Hochberg and Storey procedures. At target FDR levels $\alpha \leq 0.5$, the Benjamini-Hochberg and Storey methods are unable to make more than a few discoveries, while the accumulation tests produce many discoveries.

Comparing the accumulation tests that are studied here, the SeqStep and SeqStep+ procedures are almost identical, showing that when the number of discoveries is high, the slight correction in the definition of the SeqStep+ method (4) has essentially no loss of power relative to SeqStep. The HingeExp method yields substantially more discoveries than SeqStep and SeqStep+, which in turn yield more discoveries than ForwardStop.

[5]Data available at `http://www.ncbi.nlm.nih.gov/sites/GDSbrowser?acc=GDS2324` or via the `GEOquery` package [11] in R.

[6]For the ForwardStop and HingeExp methods, since $\binom{10}{5} = 252$, our permutation test p-values take values in the set $\{\frac{1}{252}, \frac{2}{252}, \ldots, \frac{252}{252}\}$. The possibility of $p_i = 1$ for some genes $i$ is problematic because $\mathsf{h}(1) = +\infty$ for these methods. Therefore we shift the p-values slightly to take values $\{\frac{1}{253}, \frac{2}{253}, \ldots, \frac{252}{253}\}$. Although technically this may violate the FDR control properties of the methods, the shift is extremely small and should not cause issues.

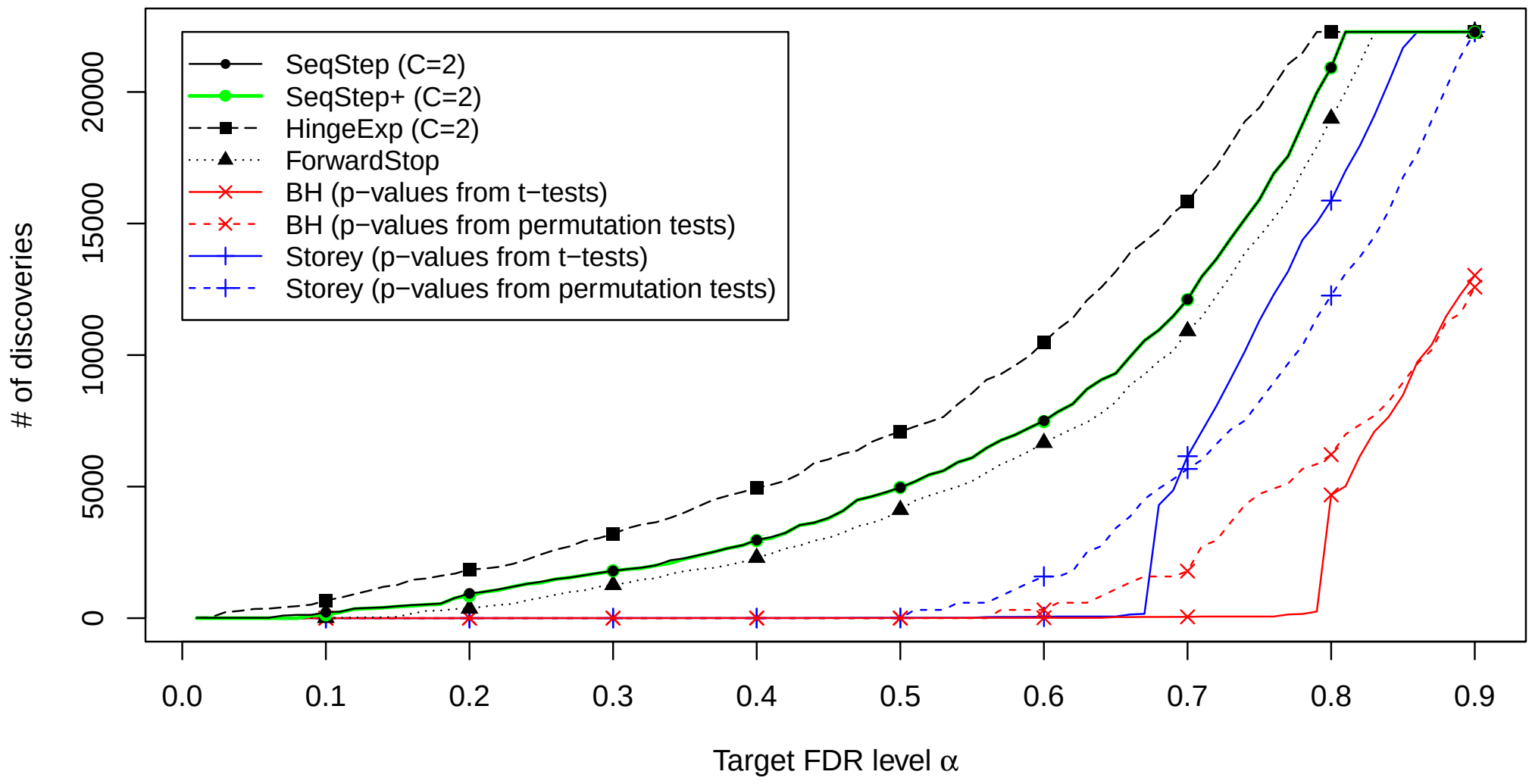

*Figure 6: Results for the differential gene expression experiment: for each method, the plot shows the number of discoveries made (i.e. the number of genes selected as showing significant change in expression at the low drug dosage), at a range of target FDR values $\alpha$. Note that the SeqStep and SeqStep+ methods are nearly indistinguishable in the plot.*

Overall, the comparisons across the different accumulation tests examined here confirms the higher power attained by HingeExp compared to existing methods in the family. We also see substantial power gain of the accumulation tests as compared to the Benjamini-Hochberg and Storey procedures, both of which do not use a sequential structure and do not make use of the high-dosage data, demonstrating the benefits of the ordered hypothesis testing approach.

# 7 Conclusion

In this paper, we have considered the multiple testing problem in the context of prior information or structure on the list of hypotheses $H_1, \ldots, H_n$ to be tested. The proposed family of accumulation tests generalizes existing methods for this ordered testing problem, and the new HingeExp method within this family gives significantly higher power than the existing tests, while maintaining control of the modified false discovery rate in a finite-sample setting, and asymptotic control of the false discovery rate. Our theoretical results prove FDR control for methods in this family in general, and examine the power properties of the tests within this family. These methods are a natural fit for any multiple testing problem where there is an inherent ordering to the hypotheses, but many other settings can be framed in this way as well—our real data experiment, which uses measurements of gene expression level across a gradient of drug dosages, indicates that we can achieve strong power gains by framing the problem as an ordered hypothesis testing problem.

In general, existing structure within the multiple testing problem can substantially increase our power to detect signals while maintaining low false positives. In addition to the ordered testing setting considered here, real problems can exhibit many types of structure—for example a hierarchical structure (represented as a tree) on the set of hypotheses, a grouped structure, or prior information that gives a prior distribution for the status (true signal or null and/or strength of the signal) of each hypothesis. These types of multiple testing problems offer more information or a different type of structure relative to the ordered testing problem considered here. Existing literature offers some methods for several of these settings. It would be of interest to examine whether a general framework can encompass all of these possible structures for the multiple testing problem, to offer a single unifying approach that is flexible enough to incorporate any type of informative structure in order to discover as many signals as possible.

## Acknowledgements

The authors thank Ryan Tibshirani for kindly providing code implementing the spacing test method for the LARS regression path.

# A Appendix: proofs

## A.1 Details for the proof of Theorem 1

In this section we show details for deriving the results of Theorem 1 from the key lemma, Lemma 3. These calculations essentially follow the proof of [1, Theorem 3] but we include them here for completeness.

*Proof of Theorem 1.* First, note that the result (9) for bounded accumulation functions is simply a special case of the general result (10), since if the accumulation function $\mathsf{h}$ is bounded by $C$ then

$$\int_{t=0}^{1} \mathsf{h}(t) \wedge C \; \mathsf{d}t = \int_{t=0}^{1} \mathsf{h}(t) \; \mathsf{d}t = 1 \; .$$

Therefore it suffices to prove (10). For the first bound in (10), treating $\mathsf{FDP}(\widehat{k}^{+C}_{\mathsf{h}})$, we have

$$\begin{aligned}
\mathbb{E}\left[\mathsf{FDP}(\widehat{k}^{+C}_{\mathsf{h}})\right] &= \mathbb{E}\left[\frac{\#\{i \le \widehat{k}^{+C}_{\mathsf{h}} : i \in \mathcal{H}_0\}}{\widehat{k}^{+C}_{\mathsf{h}}} \cdot \mathbb{1}\left\{\widehat{k}^{+C}_{\mathsf{h}} > 0\right\}\right] \\
&\le \mathbb{E}\left[\frac{1 + \#\{i \le \widehat{k}^{+C}_{\mathsf{h}} : i \in \mathcal{H}_0\}}{1 + \widehat{k}^{+C}_{\mathsf{h}}}\right] \\
&= \mathbb{E}\left[\frac{1 + \#\{i \le \widehat{k}^{+C}_{\mathsf{h}} : i \in \mathcal{H}_0\}}{C + \sum_{i=1}^{\widehat{k}^{+C}_{\mathsf{h}}} \mathsf{h}(p_i)} \cdot \frac{C + \sum_{i=1}^{\widehat{k}^{+C}_{\mathsf{h}}} \mathsf{h}(p_i)}{1 + \widehat{k}^{+C}_{\mathsf{h}}}\right] \\
&\le \alpha \cdot \mathbb{E}\left[\frac{1 + \#\{i \le \widehat{k}^{+C}_{\mathsf{h}} : i \in \mathcal{H}_0\}}{C + \sum_{i=1}^{\widehat{k}^{+C}_{\mathsf{h}}} \mathsf{h}(p_i)}\right] \text{ by definition of } \widehat{k}^{+C}_{\mathsf{h}} \\
&\le \alpha \cdot \frac{1}{\int_{t=0}^{1} \mathsf{h}(t) \wedge C \; \mathsf{d}t} \text{ by Lemma 3.}
\end{aligned}$$

Turning to the first bound in (10), treating $\mathsf{mFDP}_{C/\alpha}(\widehat{k}_{\mathsf{h}})$, we have

$$\begin{aligned}
\mathbb{E}\left[\mathsf{mFDP}_{C/\alpha}(\widehat{k}_{\mathsf{h}})\right] &= \mathbb{E}\left[\frac{\#\{i \le \widehat{k}_{\mathsf{h}} : i \in \mathcal{H}_0\}}{C/\alpha + \widehat{k}_{\mathsf{h}}}\right] \\
&= \mathbb{E}\left[\frac{\#\{i \le \widehat{k}_{\mathsf{h}} : i \in \mathcal{H}_0\}}{C + \sum_{i=1}^{\widehat{k}_{\mathsf{h}}} \mathsf{h}(p_i)} \cdot \frac{C + \sum_{i=1}^{\widehat{k}_{\mathsf{h}}} \mathsf{h}(p_i)}{C/\alpha + \widehat{k}_{\mathsf{h}}}\right] \\
&\le \mathbb{E}\left[\frac{\#\{i \le \widehat{k}_{\mathsf{h}} : i \in \mathcal{H}_0\}}{C + \sum_{i=1}^{\widehat{k}_{\mathsf{h}}} \mathsf{h}(p_i)} \cdot \frac{C + \widehat{k}_{\mathsf{h}} \cdot \alpha}{C/\alpha + \widehat{k}_{\mathsf{h}}}\right] \text{ by definition of } \widehat{k}_{\mathsf{h}} \\
&= \alpha \cdot \mathbb{E}\left[\frac{\#\{i \le \widehat{k}_{\mathsf{h}} : i \in \mathcal{H}_0\}}{C + \sum_{i=1}^{\widehat{k}_{\mathsf{h}}} \mathsf{h}(p_i)}\right] \\
&\le \alpha \cdot \frac{1}{\int_{t=0}^{1} \mathsf{h}(t) \wedge C \; \mathsf{d}t} \text{ by Lemma 3.}
\end{aligned}$$

□

## A.2 Proof of Lemma 5 (bounds on random walks)

*Proof of Lemma 5.* First, define

$$\tilde{\sigma} = \max\left\{\sigma, b\sqrt{2\log_2(4/\epsilon)}\right\} .$$

Since $\tilde{\sigma} \geq \sigma$, trivially each $X_i$ is $(\tilde{\sigma}, b)$-subexponential. Then for all $\theta \in [0, \frac{1}{b}]$ and all $i \geq 1$,

$$\mathbb{E}\left[e^{\theta X_i}\right] \leq \exp\left\{\frac{\theta^2\tilde{\sigma}^2}{2}\right\} .$$

Now taking any $t \geq 1$, and any $r > 0$ such that $\theta = \frac{r}{t\tilde{\sigma}^2} \leq \frac{1}{b}$,

$$\begin{aligned}
\mathbb{P}\left\{\sum_{i=1}^{t} X_i \geq r\right\} &= \mathbb{P}\left\{\theta\sum_{i=1}^{t} X_i \geq \theta r\right\} \leq \mathbb{P}\left\{e^{\theta\sum_{i=1}^{n} X_i} \geq e^{\theta r}\right\} \\
&\leq \mathbb{E}\left[e^{\theta\sum_{i=1}^{t} X_i}\right] \cdot e^{-\theta r} \quad \text{(Markov inequality)} \\
&\leq \exp\left\{\frac{t\theta^2\tilde{\sigma}^2}{2}\right\} \cdot e^{-\theta r} \\
&= \exp\left\{-\frac{r^2}{2t\tilde{\sigma}^2}\right\} \quad \text{by taking } \theta = \frac{r}{t\tilde{\sigma}^2} .
\end{aligned}$$

By an identical argument, the same bound holds for $\mathbb{P}\left\{\sum_{i=1}^{t} X_i \leq -r\right\}$, and therefore, for all $t \geq 1$ and $r \leq \frac{t\tilde{\sigma}^2}{b}$,

$$\mathbb{P}\left\{\left|\sum_{i=1}^{t} X_i\right| \geq r\right\} \leq 2\exp\left\{-\frac{r^2}{2t\tilde{\sigma}^2}\right\} .$$

Set $r = \sqrt{2\log_2(4/\epsilon)} \cdot \tilde{\sigma} \cdot \sqrt{t\log(1+t)}$. Since $\sqrt{t\log(1+t)} \leq t$ for all $t \geq 1$, note that the upper bound on $r$ is satisfied by definition of $\tilde{\sigma}$. Then we get

$$\mathbb{P}\left\{\left|\sum_{i=1}^{t} X_i\right| \geq \sqrt{2\log_2(4/\epsilon)} \cdot \tilde{\sigma} \cdot \sqrt{t\log(1+t)}\right\} \leq 2\exp\left\{-\log_2(4/\epsilon) \cdot \log(1+t)\right\} = 2(1+t)^{-\log_2(4/\epsilon)} .$$

Then we have

$$\begin{aligned}
&\mathbb{P}\left\{\left|\sum_{i=1}^{t} X_i\right| \leq \sqrt{2\log_2(4/\epsilon)}\tilde{\sigma}\sqrt{t\log(1+t)} \text{ for all } t \geq 1\right\} \\
&\geq 1 - \sum_{t\geq 1} \mathbb{P}\left\{\left|\sum_{i=1}^{t} X_i\right| \geq \sqrt{2\log_2(4/\epsilon)}\tilde{\sigma}\sqrt{t\log(1+t)}\right\} \\
&\geq 1 - \sum_{t\geq 1} 2(1+t)^{-\log_2(4/\epsilon)} \geq 1 - \int_{t=2}^{\infty} 2t^{-\log_2(4/\epsilon)}\,\mathsf{d}t \\
&= 1 + 2\frac{\left[t^{1-\log_2(4/\epsilon)}\right]\big|_{t=2}^{\infty}}{\log_2(4/\epsilon) - 1} = 1 - \frac{\epsilon}{\log_2\left(\frac{4}{\epsilon}\right) - 1} \geq 1 - \epsilon .
\end{aligned}$$

□

## A.3 Proof of Lemma 2 (bounded accumulation functions)

*Proof of Lemma 2.* We have

$$\begin{aligned}
\mathbb{E}\left[\mathsf{h}(p_i)\right] - \mathbb{E}\left[\mathsf{h}_0(p_i)\right] &= \int_{t=0}^{1} (\mathsf{h}(t) - \mathsf{h}_0(t)) \cdot f_i(t) \, \mathsf{d}t \\
&= \int_{t=0}^{1-1/C} (\mathsf{h}(t) - 0) \cdot f_i(t) \, \mathsf{d}t + \int_{t=1-1/C}^{1} (\mathsf{h}(t) - C) \cdot f_i(t) \, \mathsf{d}t \\
&= \int_{t=0}^{1-1/C} \mathsf{h}(t) \cdot f_i(t) \, \mathsf{d}t - \int_{t=1-1/C}^{1} (C - \mathsf{h}(t)) \cdot f_i(t) \, \mathsf{d}t \\
&\geq \int_{t=0}^{1-1/C} \mathsf{h}(t) \cdot f_i(1-1/C) \, \mathsf{d}t - \int_{t=1-1/C}^{1} (C - \mathsf{h}(t)) \cdot f_i(1-1/C) \, \mathsf{d}t \\
&= f_i(1-1/C) \cdot \left[ \int_{t=0}^{1} \mathsf{h}(t) \, \mathsf{d}t - \int_{t=1-1/C}^{1} C \, \mathsf{d}t \right] \\
&= f_i(1-1/C) \cdot [1-1] = 0 \, ,
\end{aligned}$$

where the inequality is true since $f_i$ is nonincreasing and since $\mathsf{h}(t) \geq 0$ and $C - \mathsf{h}(t) \geq 0$. Furthermore, if the inequality is not strict (i.e. is an equality), then we must have

$$\int_{t=0}^{1-1/C} \mathsf{h}(t) \cdot [f_i(t) - f_i(1-1/C)] \, \mathsf{d}t = 0 \text{ and } \int_{t=1-1/C}^{1} (C - \mathsf{h}(t)) \cdot [f_i(1-1/C) - f_i(t)] \, \mathsf{d}t = 0 \, .$$

Note that, in both integrals, the integrand is nonnegative. Therefore, in order for the integrals to equal zero, it must be true that the integrands are equal to zero almost everywhere. However, if $f_i$ is strictly decreasing then the terms in square brackets are strictly positive in both integrals (except at endpoints). Therefore, in the first integral we must have $\mathsf{h}(t) = 0$ almost everywhere over $t \in (0, 1-1/C)$, and in the second integral we must have $C - \mathsf{h}(t) = 0$ almost everywhere over $t \in (1-1/C, 1)$. In other words, $\mathsf{h}(t) = \mathsf{h}_0(t)$ almost everywhere over $t \in [0,1]$. □

## A.4 Details for proof of Theorem 3

*Proof of Theorem 3 (continued).* Here we fill in the remaining details for the proof of Theorem 3, namely, we consider the cases that $\frac{1-\alpha}{1-\mu} \geq f(0)$ or $\frac{1-\alpha}{1-\mu} \leq f(1)$.

**Case 2: $\alpha$ satisfies $\frac{1-\alpha}{1-\mu} \geq f(0)$** For this case, we will show that power tends to zero. As in the first case, it will be sufficient to show that

$$\widehat{\mathsf{FDP}}_{\mathsf{h}}(k) > \alpha \text{ for all } k > \max\{\tau_n \cdot n, \log(n)\} \, . \tag{29}$$

To prove this, take any such $k$. Then, if the event (22) holds, we apply (23) to get

$$\widehat{\mathsf{FDP}}_{\mathsf{h}}(k) > \mathsf{E}\left(\frac{k}{n}\right) - \beta_n \geq \mathsf{E}(\tau_n) - \beta_n = 1 - f(\tau_n) \cdot (1-\mu) - \beta_n \, .$$

If $f(0) = \frac{1-\alpha}{1-\mu}$ exactly, then $f(0) > 1-\alpha$ and so $f(\tau_n) \geq 1-\alpha$ for sufficiently large $n$. Therefore, applying assumption (13), $f(\tau_n) \leq f(0) - \tau_n \delta$ and then

$$\widehat{\mathsf{FDP}}_{\mathsf{h}}(k) > 1 - (f(0) - \tau_n \delta) \cdot (1-\mu) - \beta_n = \alpha \, .$$

Alternately, if $f(0) < \frac{1-\alpha}{1-\mu}$, then since $f$ is continuous, for sufficiently large $n$ we have $f(\tau_n) \leq \frac{1-\alpha-\beta_n}{1-\mu}$ and then the same bound holds.

**Case 3:** $\alpha$ **satisfies** $\frac{1-\alpha}{1-\mu} \leq f(1)$ For this case, we will show that power tends to 1. As in the previous cases, it will be sufficient to show that

$$\widehat{\mathsf{FDP}}_{\mathsf{h}}(k) < \alpha \text{ for all } k < n \cdot (1 - \tau_n) \ .$$

To prove this, take any such $k$. Then, if the event (22) holds, we apply (23) to get

$$\widehat{\mathsf{FDP}}_{\mathsf{h}}(k) < \mathsf{E}\left(\frac{k}{n}\right) + \beta_n \leq \mathsf{E}(1-\tau_n) + \beta_n = 1 - f(1-\tau_n) \cdot (1-\mu) + \beta_n \ .$$

Since $f(1) \geq \frac{1-\alpha}{1-\mu} > 1 - \alpha$, we see that $f(t) \geq 1 - \alpha$ for all $t \in [1-\tau_n, 1]$, and so $f(1-\tau_n) \geq f(1) + \tau_n \delta$ by assumption (13). Then,

$$\widehat{\mathsf{FDP}}_{\mathsf{h}}(k) < 1 - (f(1) + \tau_n\delta) \cdot (1-\mu) + \beta_n \leq \alpha \ .$$

□

## A.5 Proof of Lemma 1 (FDR control for the HingeExp function)

*Proof of Lemma 1.* First, by Jensen's inequality, drawing $E_{0,1}, E_{0,2} \stackrel{\text{iid}}{\sim} \mathsf{Exponential}(1)$ independently from the p-values $p_1, \ldots, p_n$,

$$\begin{aligned}
\mathbb{E}\left[\frac{\#\{i \leq \widehat{k}_{\mathsf{h}} : i \in \mathcal{H}_0\}}{2C\alpha^{-1} + \widehat{k}_{\mathsf{h}}}\right] &= \mathbb{E}\left[\frac{\#\{i \leq \widehat{k}_{\mathsf{h}} : i \in \mathcal{H}_0\}}{C\alpha^{-1}\mathbb{E}\left[E_{0,1} + E_{0,2} \mid p_1, \ldots, p_n\right] + \widehat{k}_{\mathsf{h}}}\right] \\
&\leq \mathbb{E}\left[\mathbb{E}\left[\frac{\#\{i \leq \widehat{k}_{\mathsf{h}} : i \in \mathcal{H}_0\}}{C\alpha^{-1}(E_{0,1} + E_{0,2}) + \widehat{k}_{\mathsf{h}}} \,\middle|\, p_1, \ldots, p_n\right]\right] \\
&= \mathbb{E}\left[\frac{\#\{i \leq \widehat{k}_{\mathsf{h}} : i \in \mathcal{H}_0\}}{C\alpha^{-1}(E_{0,1} + E_{0,2}) + \widehat{k}_{\mathsf{h}}}\right] \ .
\end{aligned}$$

Next,

$$\begin{aligned}
&\mathbb{E}\left[\frac{\#\{i \le \widehat{k}_{\mathsf{h}} : i \in \mathcal{H}_0\}}{C\alpha^{-1}(E_{0,1}+E_{0,2})+\widehat{k}_{\mathsf{h}}}\right] \\
&= \mathbb{E}\left[\frac{\#\{i \le \widehat{k}_{\mathsf{h}} : i \in \mathcal{H}_0\}}{C(E_{0,1}+E_{0,2})+\sum_{i\le \widehat{k}_{\mathsf{h}}, i\in\mathcal{H}_0}\mathsf{h}(p_i)} \cdot \frac{C(E_{0,1}+E_{0,2})+\sum_{i\le \widehat{k}_{\mathsf{h}}, i\in\mathcal{H}_0}\mathsf{h}(p_i)}{C\alpha^{-1}(E_{0,1}+E_{0,2})+\widehat{k}_{\mathsf{h}}}\right] \\
&\le \mathbb{E}\left[\frac{\#\{i \le \widehat{k}_{\mathsf{h}} : i \in \mathcal{H}_0\}}{C(E_{0,1}+E_{0,2})+\sum_{i\le \widehat{k}_{\mathsf{h}}, i\in\mathcal{H}_0}\mathsf{h}(p_i)} \cdot \frac{C(E_{0,1}+E_{0,2})+\sum_{i\le \widehat{k}_{\mathsf{h}}}\mathsf{h}(p_i)}{C\alpha^{-1}(E_{0,1}+E_{0,2})+\widehat{k}_{\mathsf{h}}}\right] \\
&= \mathbb{E}\left[\frac{\#\{i \le \widehat{k}_{\mathsf{h}} : i \in \mathcal{H}_0\}}{C(E_{0,1}+E_{0,2})+\sum_{i\le \widehat{k}_{\mathsf{h}}, i\in\mathcal{H}_0}\mathsf{h}(p_i)} \cdot \frac{C(E_{0,1}+E_{0,2})+\widehat{k}_{\mathsf{h}}\cdot\widehat{\mathsf{FDP}}_{\mathsf{h}}(\widehat{k}_{\mathsf{h}})}{C\alpha^{-1}(E_{0,1}+E_{0,2})+\widehat{k}_{\mathsf{h}}}\right] \\
&\le \mathbb{E}\left[\frac{\#\{i \le \widehat{k}_{\mathsf{h}} : i \in \mathcal{H}_0\}}{C(E_{0,1}+E_{0,2})+\sum_{i\le \widehat{k}_{\mathsf{h}}, i\in\mathcal{H}_0}\mathsf{h}(p_i)} \cdot \frac{C(E_{0,1}+E_{0,2})+\widehat{k}_{\mathsf{h}}\cdot\alpha}{C\alpha^{-1}(E_{0,1}+E_{0,2})+\widehat{k}_{\mathsf{h}}}\right] \\
&= \alpha\cdot\mathbb{E}\left[\frac{\#\{i \le \widehat{k}_{\mathsf{h}} : i \in \mathcal{H}_0\}}{C(E_{0,1}+E_{0,2})+\sum_{i\le \widehat{k}_{\mathsf{h}}, i\in\mathcal{H}_0}\mathsf{h}(p_i)}\right] .
\end{aligned}$$

Next, note that $\mathsf{h}(p_i)$ is equal in distribution to $C \cdot B_i \cdot E_i$, where $B_i \sim \mathsf{Bernoulli}(1/C)$ and $E_i \sim \mathsf{Exponential}(1)$, for all $i \in \mathcal{H}_0$. (Here we assume that the variables $B_i$ and $E_i$ are all mutually independent). Therefore we have

$$\mathbb{E}\left[\frac{\#\{i \le \widehat{k}_{\mathsf{h}} : i \in \mathcal{H}_0\}}{C\alpha^{-1}(E_{0,1}+E_{0,2})+\widehat{k}_{\mathsf{h}}}\right] \le \alpha\cdot\frac{1}{C}\cdot\mathbb{E}\left[\frac{\#\{i \le \widehat{k}_{\mathsf{h}} : i \in \mathcal{H}_0\}}{E_{0,1}+E_{0,2}+\sum_{i\le \widehat{k}_{\mathsf{h}}, i\in\mathcal{H}_0} B_i\cdot E_i}\right] \le \alpha\cdot\frac{1}{C}\cdot\mathbb{E}\left[M_{\widehat{k}_{\mathsf{h}}}\right] ,$$

where we define

$$M_k = \frac{1+\#\{i\le k : i\in\mathcal{H}_0\}}{E_{0,1}+E_{0,2}+\sum_{i\le k, i\in\mathcal{H}_0} B_i\cdot E_i} .$$

Next, we prove that $M_k$ is a supermartingale with $\mathbb{E}[M_n] \le C$, and that $\widehat{k}_{\mathsf{h}}$ is a stopping time.

Let $\mathcal{F}_k$ be the $\sigma$-algebra defined by knowing $E_{0,1}, E_{0,2}$, knowing $(B_i, E_i)$ for all $i \notin \mathcal{H}_0$, knowing $(B_i, E_i)$ for all $i > k$ with $i \in \mathcal{H}_0$, and knowing $\{(B_i, E_i) : i \le k, i \in \mathcal{H}_0\}$ (note that this is an unordered set, as before in e.g. the proof of Lemma 3). Let $\tilde{\mathcal{F}}_k$ be the $\sigma$-algebra that additionally knows $B_1, \dots, B_n$.

Now we show that $\mathbb{E}[M_k \mid \mathcal{F}_{k+1}] \le M_{k+1}$. If $k+1 \notin \mathcal{H}_0$ or if $B_{k+1} = 0$, then $M_k \le M_{k+1}$ trivially. Turning to the case where $k+1 \in \mathcal{H}_0$ and $B_{k+1} = 1$, we begin by conditioning on $B_1, ..., B_n$. In that case, we see that

$$E_{0,1} + E_{0,2} + \sum_{i\le k, i\in\mathcal{H}_0} B_i \cdot E_i$$

is a sum of $(2+\sum_{i\le k, i\in\mathcal{H}_0} B_i)$ many $\mathsf{Exponential}(1)$ variables, which $\sim \mathsf{Gamma}(2+\sum_{i\le k, i\in\mathcal{H}_0} B_i, 1)$, while

$$B_{k+1}\cdot E_{k+1}$$

is equal to another (independent) $\mathsf{Exponential}(1)$ variable, which $\sim \mathsf{Gamma}(1,1)$. Using the fact that

$$\text{If } X \sim \mathsf{Gamma}(k), Y \sim \mathsf{Gamma}(l), X \perp Y, \text{then } 1). \frac{X}{X+Y} \sim \mathsf{Beta}(k,l), 2). \frac{X}{X+Y} \perp X+Y .$$

Conditioning on $B_1, ..., B_n$,

$$\frac{E_{0,1} + E_{0,2} + \sum_{i \le k, i \in \mathcal{H}_0} B_i \cdot E_i}{E_{0,1} + E_{0,2} + \sum_{i \le k+1, i \in \mathcal{H}_0} B_i \cdot E_i} \sim \mathsf{Beta}(2 + \sum_{i \le k, i \in \mathcal{H}_0} B_i, 1)$$

And conditioning on $\tilde{\mathcal{F}}_{k+1}$ yields the same result. Now, using the fact that

$$\text{If } X \sim \mathsf{Beta}(\alpha, \beta) \text{ and } \alpha > 1 \text{ then } \mathbb{E}\left[\frac{1}{X}\right] = \frac{\alpha + \beta - 1}{\alpha - 1} \ .$$

Therefore,

$$\mathbb{E}\left[\frac{E_{0,1} + E_{0,2} + \sum_{i \le k+1, i \in \mathcal{H}_0} B_i \cdot E_i}{E_{0,1} + E_{0,2} + \sum_{i \le k, i \in \mathcal{H}_0} B_i \cdot E_i} \,\middle|\, \tilde{\mathcal{F}}_{k+1}\right] = \frac{1 + \sum_{i \le k+1, i \in \mathcal{H}_0} B_i}{1 + \sum_{i \le k, i \in \mathcal{H}_0} B_i}.$$

We then have

$$\begin{aligned}
&\mathbb{E}\left[M_k \mid \mathcal{F}_{k+1}\right] \\
&= \mathbb{E}\left[\frac{1 + \#\{i \le k : i \in \mathcal{H}_0\}}{E_{0,1} + E_{0,2} + \sum_{i \le k, i \in \mathcal{H}_0} B_i \cdot E_i} \,\middle|\, \mathcal{F}_{k+1}\right] \\
&= \mathbb{E}\left[\frac{1 + \#\{i \le k : i \in \mathcal{H}_0\}}{E_{0,1} + E_{0,2} + \sum_{i \le k+1, i \in \mathcal{H}_0} B_i \cdot E_i} \cdot \frac{E_{0,1} + E_{0,2} + \sum_{i \le k+1, i \in \mathcal{H}_0} B_i \cdot E_i}{E_{0,1} + E_{0,2} + \sum_{i \le k, i \in \mathcal{H}_0} B_i \cdot E_i} \,\middle|\, \mathcal{F}_{k+1}\right] \\
&= \mathbb{E}\left[\frac{1 + \#\{i \le k : i \in \mathcal{H}_0\}}{E_{0,1} + E_{0,2} + \sum_{i \le k+1, i \in \mathcal{H}_0} B_i \cdot E_i} \cdot \mathbb{E}\left[\frac{E_{0,1} + E_{0,2} + \sum_{i \le k+1, i \in \mathcal{H}_0} B_i \cdot E_i}{E_{0,1} + E_{0,2} + \sum_{i \le k, i \in \mathcal{H}_0} B_i \cdot E_i} \,\middle|\, \tilde{\mathcal{F}}_{k+1}\right] \,\middle|\, \mathcal{F}_{k+1}\right]
\end{aligned}$$

since here $k+1 \in \mathcal{H}_0$, $\#\{i \le k : i \in \mathcal{H}_0\} = \#\{i \le k+1 : i \in \mathcal{H}_0\} - 1$ is known, given $\tilde{\mathcal{F}}_{k+1}$

$$\begin{aligned}
&= \mathbb{E}\left[\frac{1 + \#\{i \le k : i \in \mathcal{H}_0\}}{E_{0,1} + E_{0,2} + \sum_{i \le k+1, i \in \mathcal{H}_0} B_i \cdot E_i} \cdot \frac{1 + \sum_{i \le k+1, i \in \mathcal{H}_0} B_i}{1 + \sum_{i \le k, i \in \mathcal{H}_0} B_i} \,\middle|\, \mathcal{F}_{k+1}\right] \\
&= \frac{1 + \sum_{i \le k+1, i \in \mathcal{H}_0} B_i}{E_{0,1} + E_{0,2} + \sum_{i \le k+1, i \in \mathcal{H}_0} B_i \cdot E_i} \cdot \mathbb{E}\left[\frac{1 + \#\{i \le k : i \in \mathcal{H}_0\}}{1 + \sum_{i \le k, i \in \mathcal{H}_0} B_i} \,\middle|\, \mathcal{F}_{k+1}\right] \\
&\le \frac{1 + \sum_{i \le k+1, i \in \mathcal{H}_0} B_i}{E_{0,1} + E_{0,2} + \sum_{i \le k+1, i \in \mathcal{H}_0} B_i \cdot E_i} \cdot \frac{1 + \#\{i \le k+1 : i \in \mathcal{H}_0\}}{1 + \sum_{i \le k+1, i \in \mathcal{H}_0} B_i} \\
&= \frac{1 + \#\{i \le k+1 : i \in \mathcal{H}_0\}}{E_{0,1} + E_{0,2} + \sum_{i \le k+1, i \in \mathcal{H}_0} B_i \cdot E_i} = M_{k+1} \ ,
\end{aligned}$$

where the inequality in the next-to-last step comes from Lemma 4. This proves that $M_k$ is a super-martingale. Finally, we have

$$\begin{aligned}
\mathbb{E}\left[M_n\right] &= \mathbb{E}\left[\frac{1+|\mathcal{H}_0|}{E_{0,1}+E_{0,2}+\sum_{i\in\mathcal{H}_0} B_i\cdot E_i}\right] \\
&= \mathbb{E}\left[\mathbb{E}\left[\frac{1+|\mathcal{H}_0|}{E_{0,1}+E_{0,2}+\sum_{i\in\mathcal{H}_0} B_i\cdot E_i}\;\middle|\;\sum_{i\in\mathcal{H}_0} B_i\right]\right] \\
&= \mathbb{E}\left[\mathbb{E}\left[\frac{1+|\mathcal{H}_0|}{\mathsf{Gamma}(2+\sum_{i\in\mathcal{H}_0} B_i, 1)}\;\middle|\;\sum_{i\in\mathcal{H}_0} B_i\right]\right] \\
&\leq \mathbb{E}\left[\frac{1+|\mathcal{H}_0|}{(2+\sum_{i\in\mathcal{H}_0} B_i)-1}\right] \\
&\leq C\ ,
\end{aligned}$$

where again we apply Lemma 4 for the last step. □